\documentclass[10pt,journal,compsoc]{IEEEtran}
\ifCLASSOPTIONcompsoc
  \usepackage[nocompress]{cite}
\else
  \usepackage{cite}
\fi
\ifCLASSINFOpdf
  \usepackage[pdftex]{graphicx}
\else
\fi
\usepackage{amsmath}
\usepackage{amsthm}
\usepackage{amsmath,amssymb,amsfonts}
\usepackage[ruled,linesnumbered]{algorithm2e}
\usepackage{caption}
\usepackage{multirow}
\usepackage{textcomp}
\usepackage{multicol}
\usepackage{cuted}

\usepackage{array}

\usepackage{amsmath}
\usepackage{bbm}
\begin{document}
%
\title{Robust Scheduling in Cloud Environment Based on Heuristic Optimization Algorithm}
%
%
%
%

\author{Jiaxin~Zhou, Siyi~Chen*, Haiyang~Kuang, Xu Wang

\IEEEcompsocitemizethanks{\IEEEcompsocthanksitem Siyi Chen is with the School of Automation and Industrial Electronics, Xiangtan University, Hunan, China, 411105.\protect\\
E-mail: c.siyi@xtu.edu.cn.
\IEEEcompsocthanksitem Jiaxin Zhou, Haiyang Kuang are with Xiangtan University.\protect}
\thanks{Manuscript received April 19, 2005; revised August 26, 2015.}}

%
%

\markboth{Journal of \LaTeX\ Class Files,~Vol.~14, No.~8, August~2015}%
{Shell \MakeLowercase{\textit{et al.}}: Customer-Satisfaction-Aware Profit Maximization Problem in Cloud Computing with Deadline Constraint}
%



\IEEEtitleabstractindextext{%
\begin{abstract}
Aiming at analyzing performance in cloud computing, some unpredictable perturbations which may lead to performance downgrade are essential factors that should not be neglected. To avoid performance downgrade in cloud computing system, it is reasonable to measure the impact of the perturbations, and further propose a robust scheduling strategy to maintain the performance of the system at an acceptable level. In this paper, we first describe the supply-demand relationship of service between cloud service providers and customers, in which the profit and waiting time are objectives they most concerned. Then, on the basis of introducing the lowest acceptable profit and longest acceptable waiting time for cloud service providers and customers respectively, we define a robustness metric method to declare that the number and speed of servers should be adequately configured in a feasible region, such that the performance of cloud computing system can stay at an acceptable level when it is subject to the perturbations. Subsequently, we discuss the robustness metric method in several cases, and propose heuristic optimization algorithm to enhance the robustness of the system as much as possible. At last, the performances of the proposed algorithm are validated by comparing with DE and PSO algorithm, the results show the superiority of the proposed algorithm.
\end{abstract}

}

\maketitle

\IEEEdisplaynontitleabstractindextext

%
\IEEEpeerreviewmaketitle

\IEEEraisesectionheading{\section{Introduction}\label{sec:introduction}}

%
%
%
%
\IEEEPARstart{T}{oday}, cloud computing has developed into an integral part of current and future information technology. Cloud services are widely used by many organizations and individual users due to the presence of various advantages such as high efficiency, reliability and low cost \cite{jayaprakash2021systematic}.~However, with the dramatic increase in cloud usage, cloud service providers are unable to provide unlimited resources for customers to cope with surging or fluctuating demands \cite{li2012online}, so cloud service providers need to improve their quality of service (QoS). Improving QoS means configuring higher service-level agreement (SLA) standards. This means that service fees, revenues and consumption costs become higher. Therefore, pursuing a high expectation of SLA is a double-edged sword, which means that the increase in revenue won’t necessarily reduce expenses. However, configuring SLA low standards can have a similar situation. For this reason it is essential to configure the parameters wisely in the cloud environment so that users’ needs and expectations of cloud service providers are largely met.

Resource allocation in cloud computing involves scheduling and provision of resources, meanwhile, available infrastructure \cite{alam2022cloud}, service level agreements \cite{mahmud2020profit}, cost  \cite{li2020resource} and energy factors \cite{albarracin2023exploration} are also taken into consideration. However, when it comes to resource allocation, there will always be some perturbations in the current cloud environment. For these perturbations we can reasonably make the assumption. First, when there are fewer processes, the CPU cores are not working at full capacity in their normal state, and they may break down after running for a long enough time. Second, if an unexpectedly high load arrives, it will cause other cores that are dormant to be woken up to provide a more efficient service. Alternatively, if the bandwidth is too small, it may cause the service to become inefficient. The first scenario means that the number of service requests completed by the cloud platform per unit time will decrease, which will result in a decrease in the revenue of cloud service providers and an increase in the waiting time of customers. In fact, these can also be applied to the third scenario. In addition, the second scenario means increase the overhead of the cloud platform. With the above considerations, we can know that the benefit of cloud service providers and customers will be injured. Once the profit falls below the threshold, the normal operation of the system is affected. When the waiting time exceeds the threshold, customers will not come back after experiencing the current unsatisfactory service. In conclusion, changes in configuration parameters under the influence of perturbations will affect the response performance of the system and may even cause the corresponding performance index to exceed the given threshold in extreme cases. This shows that the perturbation factor should be fully considered when configuring the cloud platform. Therefore, on the premise of the introduction of disturbances, making response characteristics meet the desired performance specifications can be classified as a study of the robustness of the system. On this basis, three questions are elicited. First, how to model the cloud environment with perturbations? Second, how to perform robustness metrics on it? Third, how to obtain the optimal configuration to ensure robustness?

There is a lot of discussion about perturbations factors in this problem of resource scheduling. For example, Gong et al. \cite{gong2019adaptive} proposed an adaptive control method for resource allocation to respond adaptively to dynamic requests for workload and resource requirements. Multiple resources are allocated to multiple services based on dynamically fluctuating requests, and interference between co-hosted services is considered to ensure QoS in case of insufficient resource pools. 
Hui et al. \cite{hui2019new} investigated a resource allocation mechanism based on a deterministic differential equation model and extended it to a MEC network environment with stochastic perturbations to develop a new stochastic differential equation model. The relationship between the oscillation and the intensity of the stochastic perturbation is quantitatively analyzed. 
Zhang et al. \cite{chen2016robust} propose an efficient stochastic auction mechanism that takes into account the presence of perturbations and new applications based on smoothing analysis and stochastic parsimony for dynamic VM provisioning and pricing in geographically distributed cloud data centers.

For the perspective of robustness,it is a very extensive topic in control field. For example, Chen et al. \cite{chen2016robust} proposed a robust nonlinear controller in order to study the robustness of the controller to model uncertainties and external perturbations. The stability of the closed-loop system is ensured by gradually stabilizing each subsystem. Hua et al. \cite{hua2016robust} studied the robust controller design problem for a class of fractional-order nonlinear systems with time-varying delays. A reduced-order observer and an output feedback controller are designed. By selecting the appropriate function, it is proved that the designed controller can make the fractional order system asymptotically stable. Wang et al. \cite{wang2019robust} solved the problem of robust adaptive finite-time tracking control for a class of mechanical systems in the presence of model uncertainty, unknown external disturbances and input nonlinearity including saturation and dead zone. An adaptive tracking control scheme is proposed by using recursive design method. As mentioned earlier, robustness has yielded excellent results in control fields. While for the research in cloud environment, for the best of our knowledge, few studies are found.~
X.Ye et al.~\cite{ye2017energy} established a multi-objective virtual machine layout model for the energy-saving virtual machine placement problem, considering the various needs of cloud providers and users. An energy-efficient algorithm is proposed to minimize energy consumption and improve load balancing, resource utilization and robustness. M. Smara et al. \cite{smara2022robustness} propose a new formalized framework for using the DRB (distributed recovery block) scheme to build reliable and usable cloud components. The cloud reliability is improved by building fault shielding nodes to uniformly handle software and hardware failures.~ N.S.V. Rao et al.~\cite{rao2012cloud} proposed a game theory approach to provide and operate infrastructure under a unified cost model. The conditions for strengthening infrastructure are derived to achieve a higher level of robustness.

In order to solve the problem of unreasonable resource allocation in cloud environment, there has been a lot of discussion on it. For example, Li et al. \cite{li2010adaptive} proposed an adaptive resource allocation algorithm for cloud systems with preemptable tasks, which adaptively adjusts resource allocation based on updates of actual task execution and improves cloud utilization.~M.Ficco et al.~\cite{ficco2018coral} proposed a meta-heuristic approach for cloud resource allocation, which is based on a bionic reef optimization paradigm to model cloud elasticity in cloud data centers and on classical game theory to optimize resource reallocation patterns for cloud providers' optimization goals as well as for customers' needs. J. Praveenchandar et al. \cite{praveenchandar2021dynamic} proposed an improved task scheduling and optimal power minimization method for efficient dynamic resource allocation process. By using the prediction mechanism and the dynamic resource table update algorithm, the resource allocation efficiency in task completion and response time is realized. P. Devarasetty et al. \cite{devarasetty2021genetic} proposed an improved resource allocation optimization algorithm considering the objectives of minimizing deployment cost and improving QoS performance.~Different customer QoS requirements are considered and resources are allocated within a given budget. P. Durgadevi et al. \cite{durgadevi2020resource} proposed a hybrid optimization algorithm where request speed and size are evaluated and used to allocate resources at the server side in a given system.

In this paper, we propose a robust scheduling scheme for a class of specific cloud platform with some unpredictable perturbations being taken into consideration. During the analysis of such scheme, the profit of cloud service providers and the waiting time of customers will be affected by the perturbations in the number and speed of the servers. Under the premise of ensuring the robustness, appropriate scheduling algorithms are constructed to optimize the profit and waiting time as much as possible. The main contributions of this paper are summarized as follows:
\begin{itemize}
  \item Design the boundedness constraints for the variations of the size and speed of servers impacted by unpredictable perturbations in cloud computing system, which reflects the worst cases that cloud service providers and customers accept.
  \item Propose robustness analysis method to satisfy the requirements of cloud service providers and customers as much as possible, in which a concept of robustness radius is defined to describe the extreme situation that the worst cases for cloud service providers and customers can be optimized.
  \item On the basis of adopting meta-heuristic algorithm and successive approximation approach, investigate a combinatorial optimization method to obtain the optimal configuration of the size and speed of servers in the presence of unpredictable perturbations.
  \item Perform a series of experimental comparisons to validate the superiority of the proposed algorithm.

\end{itemize}

The remainder of this paper is organized as follows. Section \ref{sec: related work} provides a review of related work. Section \ref{sec: the models}  builds the system framework as well as the revenue model and cost model. Section \ref{sec: problem description} describes the problems, defines the robustness metric. Section \ref{sec: Robustness analysis} proposes the optimal resource scheduling method considering robustness, performs the analysis of system robustness, and analyzes the performance of the algorithm.

\section{Related Work}\label{sec: related work}
In this section, we first review the recent work related to the problem of profit maximization by resource provisioning, and then review the work related to optimize the waiting time in cloud systems. 

When discussing the profit maximization problem in cloud computing, some researches are investigated under specific constraints. For example, C.K. Swain et al.\cite{swain2020constraint} proposed the constraint-aware profit-ranking and mapping heuristic algorithm  (HOM-APM) for efficient scheduling of tasks to obtain maximum profit. Cong et al. \cite{cong2022multiserver} proposed an efficient grouped GWO-based heuristic method, which determined the optimal multi-server configuration for a given customer who demand to maximize the profitability of cloud services. Wang et al. \cite{wang2023efficient} presented an resource collaboration and expanding scheme ,which maximized long-term profitability and maintainied computational latency.  Chen et al. \cite{chen2022customer} analyzed the deadline constrained profit maximization problem in a cyclic cascaded queueing system. Regardless of the constraints, some researches have also been taken into consideration. For example, M. Najm et al.\cite{najm2020vm} proposed an algorithm for migrating virtual machines between data centers in a federated cloud, through while the operational costs was reduced of cloud providers. P. Pradhan \cite{pradhan2016modified} proposed a modified round robin resource allocation algorithm to satisfy customer demands by reducing the waiting time. Du et al.\cite{du2019learning} used a black-box approach and leveraged model-free deep reinforcement learning (DRL), through while the dynamics of cloud subscribers was captured and the cloud provider's net profit was maximized.  All of these researches can effectively help us choose the appropriate method to dicuss the profit maximization problem.\\
\indent
Apart from the profit maximization problem, customer waiting time, as an important issue reflecting the efficiency of the service and the customers attitude, simultaneously, affecting the level of customers satisfaction with the quantity of service, is also worthy discussing. For example, Zhang et al. \cite{zhang2011dynamic}  used Model Predictive Control (MPC) to find the best match for customer demand issues in terms of supply and price, as well as to minimize the average wait time for requests. A. Belgacem et al. \cite{belgacem2022intelligent} proposed a new intelligent multi-agent system and reinforcement learning method (IMARM), which made the system with reasonable load balance as well as execution time by moving the virtual machines to the optimal state. S. Saha et al. \cite{saha2016novel} introduced a genetic algorithm based task scheduling algorithm,  which used queuing model to minimize the waiting time and queue length of the system. E.S. Alkayal, et al. \cite{alkayal2016efficient} developed an algorithm based on a new sequencing policy that schedules tasks to virtual machines, such that waiting time can be minimized while system throughput can be maximized S. Pal et al. \cite{pal2016adaptation} provided optimal sequences by designing a system with Johnson scheduling algorithm. The job scheduling model was improved by reducing the waiting time and queue length using a multi-server finite capacity queuing model. A.V. Krishna et al. \cite{vijaya2020task} used and implemented a mixture of shortest job priority and scheduling based priority in a cloud environment. The average waiting time and turnaround time were greatly reduced. Besides, the efficiency of cloud resource management was significantly improved. According to the supply-demand relationship between cloud service providers and customers, profit maximization and the waiting time are the primary concerns of cloud service providers and customers, respectively. For this reason, in our robustness discussion for cloud environments where interference is present, these two aspects are considered simultaneously.

\section{The models}\label{sec: the models}

\subsection{Multi-server system}
\begin{figure}
  \centering
  \includegraphics[width=0.53\textwidth]{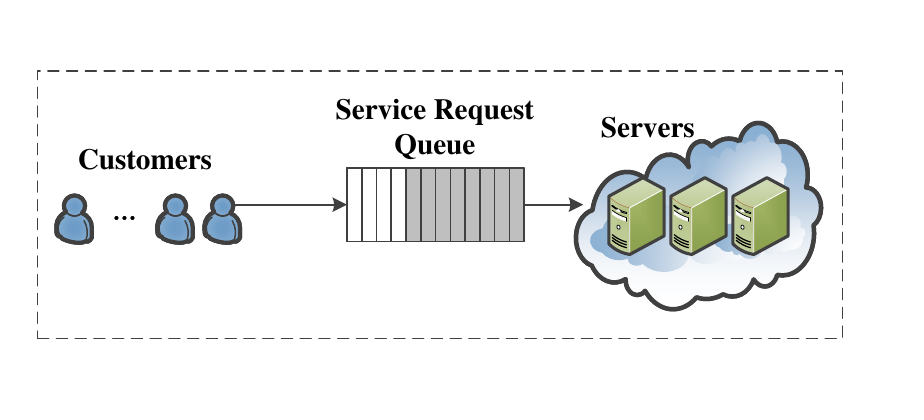}
  \caption{Multi-server queuing system.}
  \label{Fig:xit.pdf}
\end{figure}

In this paper, we consider a M/M/m multi-server queuing system, which is shown in Figure. \ref{Fig:xit.pdf}. For the multi-server system, it has $m$ identical servers with speed $s$ (hundred billion instructions per hour). When a customer arrives the queue with service requests, it will be served immediately as long as any server is available. Due to the randomness of the arrival of the customers, we consider the time it takes customers to arrive at the system as a random variable with an independent and identically distributed (i.i.d.) exponential distribution whose mean is $1/\lambda $ sec. In other words, the service requests follow a Poisson process with arrival rate $\lambda$. The task execution demand is measured by the number of instructions, which can be denoted by a the exponential random variable $r$ with mean $\overline r $. Thus, the execution time of a task on the multi-server system can also be considered as an exponential random variable $t = r/s$ with mean $\overline t  = \overline r /s$.~ Moreover, the service rate can be denoted as $\mu {\text{ }} = {\text{ }}1/\overline t {\text{ }} = {\text{ }}s/\overline r $ and the utilization factor is defined as $\rho {\text{ }} = \lambda /m\mu {\text{ }} = {\text{ }}\lambda /m \times \overline r /s$, which means the percentage of the average time when the server is in busy state. Let ${p_k}$ denotes the probability that there are $k$ service requests (waiting or being processed) in the M/M/m queuing system. Then, we have

\begin{equation}\label{Eqs:Eqs1}
   {p_k} = \left\{ {\begin{array}{*{20}{c}}
  {{p_0}\frac{{{{\left( {m\rho } \right)}^k}}}{{k!}},}&{k < m} \\ 
  \\
  {{p_0}\frac{{{m^m}{\rho ^k}}}{{m!}},}&{k \geqslant m} 
\end{array}} \right.
\end{equation}
where
\begin{equation}\label{Eqs:Eqs2}
  {p_0}{\text{ }} = {\left( {\sum\limits_{k = 0}^{m - 1} {\frac{{{{(m\rho )}^k}}}{{k!}} + \frac{{{{(m\rho )}^m}}}{{m!}}\frac{{1{\text{ }}}}{{1 - \rho }}} {\text{ }}} \right)^{ - 1}}\
\end{equation}

To ensure the ergodicity of queue system, it is obviously that the condition $0 < \rho  < 1$ should be satisfied.

On this basis, when all servers in a multi-server system are occupied by executing service requests, newly arrived service requests must wait in the queue. We express the probability (i.e., the probability that a newly arrived service request have to wait) as follows.

\begin{equation}\label{Eqs:Eqs3}
  {p_q} = \sum\limits_{k = m}^\infty  {{p_k} = \frac{{{p_{m}}}}{{1 - \rho }} = {p_0}\frac{{{{\left( {m\rho } \right)}^m}}}{{m!}} \cdot \frac{1}{{1 - \rho }}} 
\end{equation}

\subsection{Waiting time distribution}
When enjoying the services, customers always want to have shorter waiting time. Therefore, it is necessary to model the waiting time of customers, such that it can be better quantified. With this premise in mind, we denote $W$ as the waiting time of the service request arriving at the multi-server system. Notice that, the probability distribution function (pdf) of waiting time in M/M/m queuing system have already been fully considered in previous literatures, which is shown as follow \cite{cao2012optimal}. 
\begin{equation}\label{Eqs:Eqs4}
  f_W\left( t \right) = \left( {1 - {p_q}} \right)u\left( t \right) + m\mu {p_m}{e^{ - \left( {1 - \rho } \right)m\mu t}}
\end{equation}
where  ${p_m}= {p_0}{{{\left( {m\rho } \right)}^m}}/{{m!}}$ and $u\left( t \right)$ is unit impulse function, which is defined as
\begin{equation}\label{Eqs:Eqs5}
  {u_z}\left( t \right) = \left\{ {\begin{array}{*{20}{c}}
  {z,}&{0 \leqslant t \leqslant \frac{1}{z}} \\ 
  {0,}&{t > \frac{1}{z}} 
\end{array}} \right.
\end{equation}

Let $z \to \infty $, then we have 
\begin{equation}\label{Eqs:Eqs6}
  u\left( t \right) = \mathop {\lim }\limits_{z \to \infty }^{} {u_z}\left( t \right)
\end{equation}

The function ${u_z}\left( t \right)$ has the following properties
\begin{equation}\label{Eqs:Eqs7}
  \int_0^\infty  {{u_z}} \left( t \right)dt = 1
\end{equation}
and
\begin{equation}\label{Eqs:Eqs8}
  \int_0^\infty  {{u_z}} \left( t \right)dt = z\int_0^{1/z} {tdt = \frac{1}{{2z}}} 
\end{equation}

Further, according to \eqref{Eqs:Eqs4} and \eqref{Eqs:Eqs7}, we can calculate the expectation of waiting time as follow.
\begin{equation}\label{Eqs:Eqs9}
  \begin{array}{*{20}{l}}
  T&{ = \int\limits_{ 0}^{ + \infty } {t \cdot f_W\left( t \right)dt} } \\ 
  {}&{ = \int\limits_{ 0 }^{ + \infty } {t \cdot \left( {\left( {1 - {p_q}} \right)u\left( t \right) + m\mu {p_m}{e^{ - \left( {1 - \rho } \right)m\mu t}}} \right)dt} } \\ 
  {}&{ = \int\limits_{ 0 }^{ + \infty } {t \cdot m\mu {p_m}{e^{ - \left( {1 - \rho } \right)m\mu t}}dt} } \\ 
  {}&{ = \frac{{{p_m}}}{{m\mu {{\left( {1 - \rho } \right)}^2}}}} 
\end{array}
\end{equation}

\subsection{Profit modeling}
Cloud service providers play an important role in maintaining the cloud service platform as well as serving customer requests. During the operation of the platform, for the cloud service provider, revenue can be obtained from the customers whose requests have to be served, while operational costs have to be paid for maintaining the platform. In this paper, we denote $\varepsilon$ and $C$ as the revenue and the cost of cloud service provider respectively. Substracting the former to the latter, the profit of cloud service provider can be obtained, which can be demonstrated in  \eqref{Eqs:Eqs10}.
\begin{equation}\label{Eqs:Eqs10}
  {G} = \varepsilon  - C
\end{equation}
where, $G$ is the profit obtained by cloud serivce provider. In the next subsection, the revenue and cost of cloud service provider will be modelled in detail.

\subsubsection{Revenue modeling}
The customers have to pay for the services provided by the cloud service provider. To evaluate the relationship between the quality of services and the corresponding charges to the customers, SLA is adopted. In this paper, we use flat rate pricing for service requests because it is relatively intuitive and easy to be obtained. Based on this, the service charge function $R$ can be expressed as
\begin{equation}\label{Eqs:Eqs11}
  R(r,W) = \left\{ {\begin{array}{*{20}{c}}
  {{a}r,}&{0 \leqslant W \leqslant D} \\ 
  {0,}&{W > D} 
\end{array}} \right.
\end{equation}

where ${a}$ is a constant that represents the fee per unit of service, $D$ is the maximum tolerable time that a service request can wait (i.e., the deadline). 
In this paper,  when the waiting time does not exceed the maximum value, the fee that the customer will pay is considered as a constant value. Otherwise, the customer will enjoy the services for free. 

Based on \eqref{Eqs:Eqs4} and \eqref{Eqs:Eqs11}, we can calculate the expectations of $R(r, W)$

\begin{equation}\label{Eqs:Eqs12}
\begin{aligned}
R(r) & =H\left(R\left(r, W\right)\right) \\
& =\int_0^{\infty} f_{W}(t) a r d t \\
& =a r \int_0^{\infty}\left[\left(1-P_{q}\right) u(t)+m \mu p_{m} e^{-\left(1-\rho\right) m \mu t}\right] d t \\
& =a r \int_0^D\left[\left(1-P_{q}\right) u(t)+m \mu p_{m} e^{-\left(1-\rho\right) m \mu t}\right] d t \\
& =a r\left[\left(1-P_{q}\right)-\frac{p_{m}}{1-\rho}\left(e^{-\left(1-\rho\right) m \mu D}-1\right)\right] \\
& =a r\left(1-\frac{p_{m}}{1-\rho} e^{-\left(1-\rho\right) m \mu D}\right)
\end{aligned}
\end{equation}

Notice that, the task execution requirement $r$ is also a random variable, which follows the exponential distribution. On this basis, the expected charge of a service request in multi-server system can be calculated as follow.
\begin{equation}\label{Eqs:Eqs13}
\begin{aligned}
\bar{R} & =H\left(R(r)\right) \\
& =\int_0^{\infty} \frac{1}{\bar{r}} e^{-z / \bar{r}} R(z) d z \\
& =\frac{a}{\bar{r}}\left(1-\frac{p_{m_1}}{1-\rho} e^{-\left(1-\rho\right) m \mu D}\right) \int_0^{\infty} e^{-z / \bar{r}} z d z \\
& =a \bar{r}\left(1-\frac{p_{m}}{1-\rho} e^{-\left(1-\rho\right) m \mu D}\right)
\end{aligned}
\end{equation}

Further, we can also obtain
\begin{equation}\label{Eqs:Eqs14}
\begin{aligned}
F_W\left( D \right)=1-\frac{p_{m}}{1-\rho} e^{-\left(1-\rho\right) m \mu D}
\end{aligned}
\end{equation}

Since the number of service requests arriving the multi-server system is $\lambda$ per unit of time, the total revenue of cloud service provider is $\lambda{a}\overline r $ if all the service requests could be served before the deadline. However, if parts of the service requests are served for free, the actual revenue earned by the cloud service provider can be described as
\begin{equation}\label{Eqs:Eqs15}
  \varepsilon  = \lambda F_W\left( D \right){a}\overline r 
\end{equation}
\subsubsection{Cost modeling}

The costs to the cloud service provider consists of two main components, namely the cost of infrastructure leasing and energy consumption. The infrastructure provider maintains a large number of servers for loan, and the cloud service provider rents the servers on request and pays the corresponding leasing fees. Assume that the rental price of one server per unit of time is $\beta $, then the server rental price of $m$ multi-server system is ${{m}}\beta $.

The cost of energy consumption, another part of the service provider's cost, consists of the price of electricity and the amount of energy consumed. In this paper, the following dynamic power model is used to express the energy consumption, which has been discussed in much of the literature \cite{chandrakasan1992low}\cite{mei2013energy}\cite{zhai2004theoretical}.
\begin{equation}\label{Eqs:Eqs16}
  {P_d} = {N_{sw}}{C_L}{V^2}f
\end{equation}
where ${N_{sw}}$ is the average gate switching factor per clock cycle, ${C_L}$ is the load capacitance, $V$ is the supply voltage, and $f$ is the clock frequency.

In the ideal case, the relationship between the supply voltage $V$ and the clock frequency $f$ can be described as $V \propto {f^\phi }$$\left( {0 < \phi  \le 1} \right)$. For some constant $s$, the execution speed of the server is linearly related to the clock frequency $s \propto f$. Therefore, the dynamic power model can be converted to ${P_d} \propto b{N_{sw}}{C_L}{s^{2\nu  + 1}}$. For the sake of simplicity, we can assume ${P_d} = b{N_{sw}}{C_L}{s^{2\nu  + 1}} = \xi {s^\alpha }$, where  $\xi  = b{N_{sw}}{C_L}$ and $\alpha  = 2\phi  + 1$. In this paper, we set ${N_{sw}}{C_L} = 4$ , $b = 0.5$, $\nu {\rm{ = }}0.55$. Hence, $\alpha {\rm{ = }}2.1$ and $\xi {\rm{ = }}2$. In addition to dynamic power consumption, each server also consumes a certain amount of static power consumption ${P^*}$.

Due to the server utilization $\rho $ affects the dynamic power consumption of the server, the average amount of energy consumption per unit time is $P =  \rho \xi {s^\alpha } + {P^*}$. Assuming an electricity price of $\delta $ per watt, the total cost per unit time for the cloud service provider can be described as
\begin{equation}\label{Eqs:Eqs17}
  C = m(\beta  + \delta ( \rho \xi {s^\alpha } + {P^*}))
\end{equation}

\section{Problem description}\label{sec: problem description}

Considering the demands of cloud service providers and customers described in the previous section, most researches have focused on achieving optimal results under certain conditions. However, to the best of our knowledge, perturbation factors have never been discussed in the modeling and solving process of this problem, which may result in some undesirable impacts for cloud service providers and customers. In this section, we will describe the impacts of specific perturbation factors on the profit of cloud service providers and the waiting time of customers, and then design a robust strategy to protect the demands of cloud service providers and customers as much as possible.

According to \eqref{Eqs:Eqs9} and \eqref{Eqs:Eqs10}, because of the sophisticated structure of $p_m$, the waiting time of customers and the profit of cloud service providers are difficult to analyzed numerically. For the sake of simplicity, the following approximations are adopted, i.e., $\sum\limits_{k = 0}^{m - 1} {\frac{{{{\left( {m\rho } \right)}^k}}}{{k!}}}  \approx {e^{m\rho }}$ and $m! \approx \sqrt {2\pi m} {\left( {\frac{m}{e}} \right)^m}$. On this basis, $p_m$ can be rewritten as follow.

\begin{equation}\label{Eqs:Eqs18}
  {p_m} \approx \frac{{1 - \rho }}{{\sqrt {2\pi m} \left( {1 - \rho } \right){{\left( {{e^\rho }/e\rho } \right)}^m} + 1}}
\end{equation}

Then, substituting \eqref{Eqs:Eqs18} into \eqref{Eqs:Eqs14}, we have.

\begin{equation}\label{Eqs:Eqs19}
  F_W\left( D \right) \approx 1 - \frac{{{e^{ - m{\mu }\left( {1 - \rho } \right)D}}}}{{\sqrt {2\pi m} \left( {1 - \rho } \right){{\left( {{{{e^\rho }} \mathord{\left/
 {\vphantom {{{e^\rho }} {e\rho }}} \right.
 \kern-\nulldelimiterspace} {e\rho }}} \right)}^m} + 1}}
\end{equation}

Further, according to \eqref{Eqs:Eqs18} and \eqref{Eqs:Eqs19}, we can obtain the approximations of the waiting time of customers and the profit of cloud service providers as follow.

\begin{equation}\label{Eqs:Eqs20}
  T = \frac{{{{\left( {\lambda e} \right)}^m}}}{{{{\left( {sm} \right)}^{m - 1}}{{\left( {sm - \lambda } \right)}^2}{e^{\frac{\lambda }{s}}}\sqrt {2\pi m}  + \left( {sm - \lambda } \right){{\left( {e\lambda } \right)}^m}}}
\end{equation}

and,

\begin{equation}\label{Eqs:Eqs21}
\begin{array}{*{20}{l}}
{G}&{ = \lambda a\overline r \left[ {1 - \frac{{{e^{ - m{\mu }\left( {1 - \rho } \right)D}}}}{{\sqrt {2\pi m} \left( {1 - \rho } \right){{\left( {{{{e^\rho }} \mathord{\left/
 {\vphantom {{{e^\rho }} {e\rho }}} \right.
 \kern-\nulldelimiterspace} {e\rho }}} \right)}^m} + 1}}} \right]}\\ 
 \\
 &- m(\beta  + \delta ( \rho \xi {s^\alpha } + {P^*}))
\end{array}
\end{equation}

In this paper, we consider the perturbation factors as the unpredictable variations in the size and speed of the servers. Notice that, the utilization factor $\rho$ in \eqref{Eqs:Eqs21} can be expressed by $\rho = \frac{\lambda\bar{r}}{ms}$. For the purpose of describing the impact of the size and speed of the servers on the profit of cloud service providers more clearly, by substituting the expression of $\rho$ into \eqref{Eqs:Eqs21}, and then setting $\bar{r}$ to 1 hundred billion instructions, we have. 

\begin{equation}\label{Eqs:Eqs22}
\begin{aligned}
{G}&={\lambda a\left[ {1 - \tfrac{{{e^{\left( { - sm + \lambda } \right)D}}{{\left( {e\lambda } \right)}^m}}}{{\sqrt {2\pi m} \left( {sm - \lambda } \right){e^{\lambda /s}}{{\left( {sm} \right)}^{m + 1}} + {{\left( {e\lambda } \right)}^m}}}} \right]}\\
  &-\left( {m\beta  + \delta\lambda\xi s^{\alpha-1} + m\xi{P^*}}) \right)
  \end{aligned}
\end{equation}

Considering \eqref{Eqs:Eqs20} and \eqref{Eqs:Eqs22}, we can express these formulas by $T = f\left( {m,s} \right)$ and $G = g\left( {m,s} \right)$ respectively. Obviously, \eqref{Eqs:Eqs20} and \eqref{Eqs:Eqs22} show the strong nonlinear characteristics, which will result in the difficulty in numerical analysis. In this case, we choose to draw the graphics of $T = f\left( {m,s} \right)$ and $G = g\left( {m,s} \right)$ to show the functional relationship intuitively. 
\begin{figure}[!t]\centering
  \centerline{\includegraphics[width=0.53\textwidth]{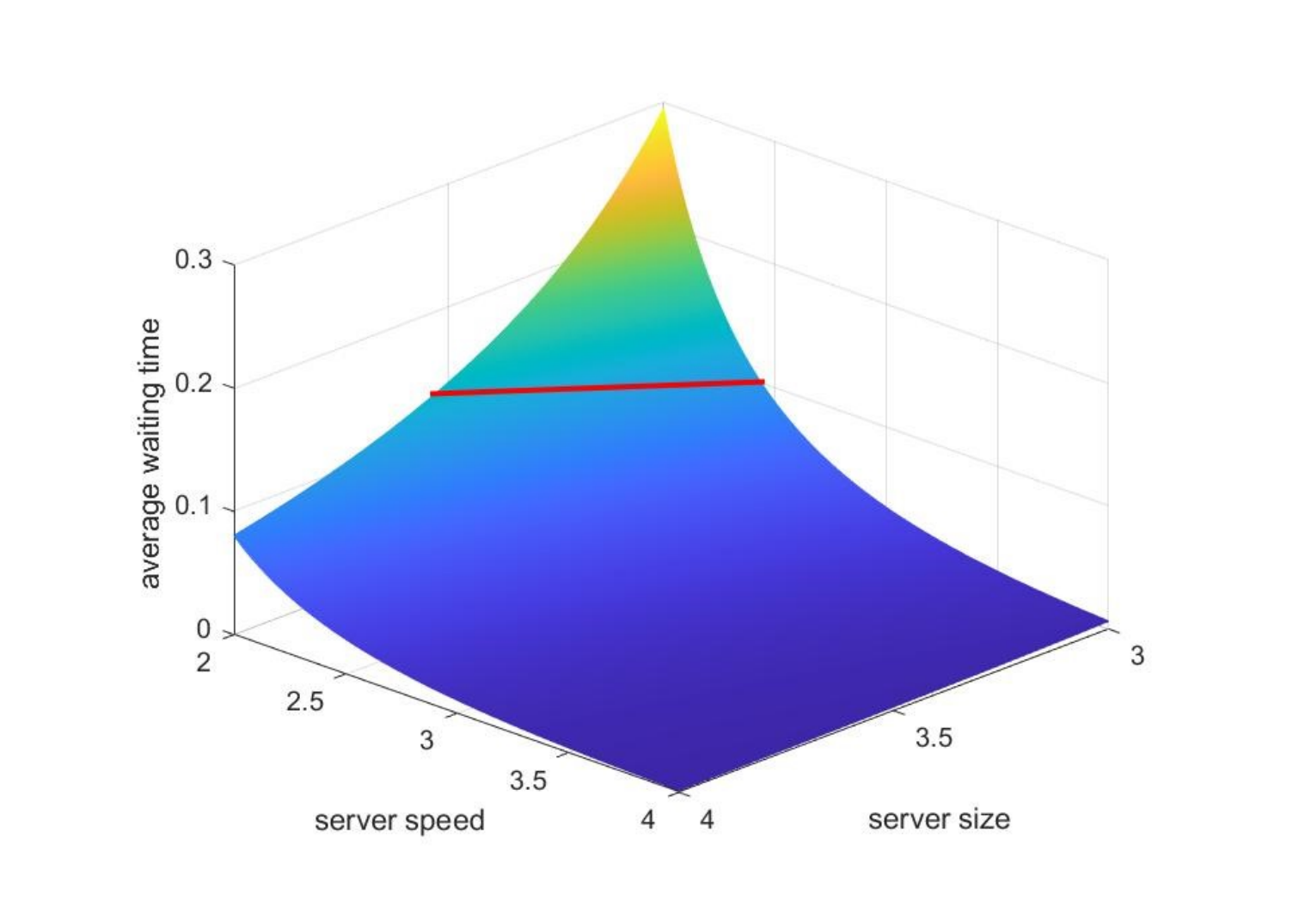} }
  \caption{The mesh of average waiting time $T$ versus $s$ and $m$.}
 \label{Fig:Fig2}
\end{figure}

\begin{figure}[!t]\centering
  \centerline{\includegraphics[width=0.53\textwidth]{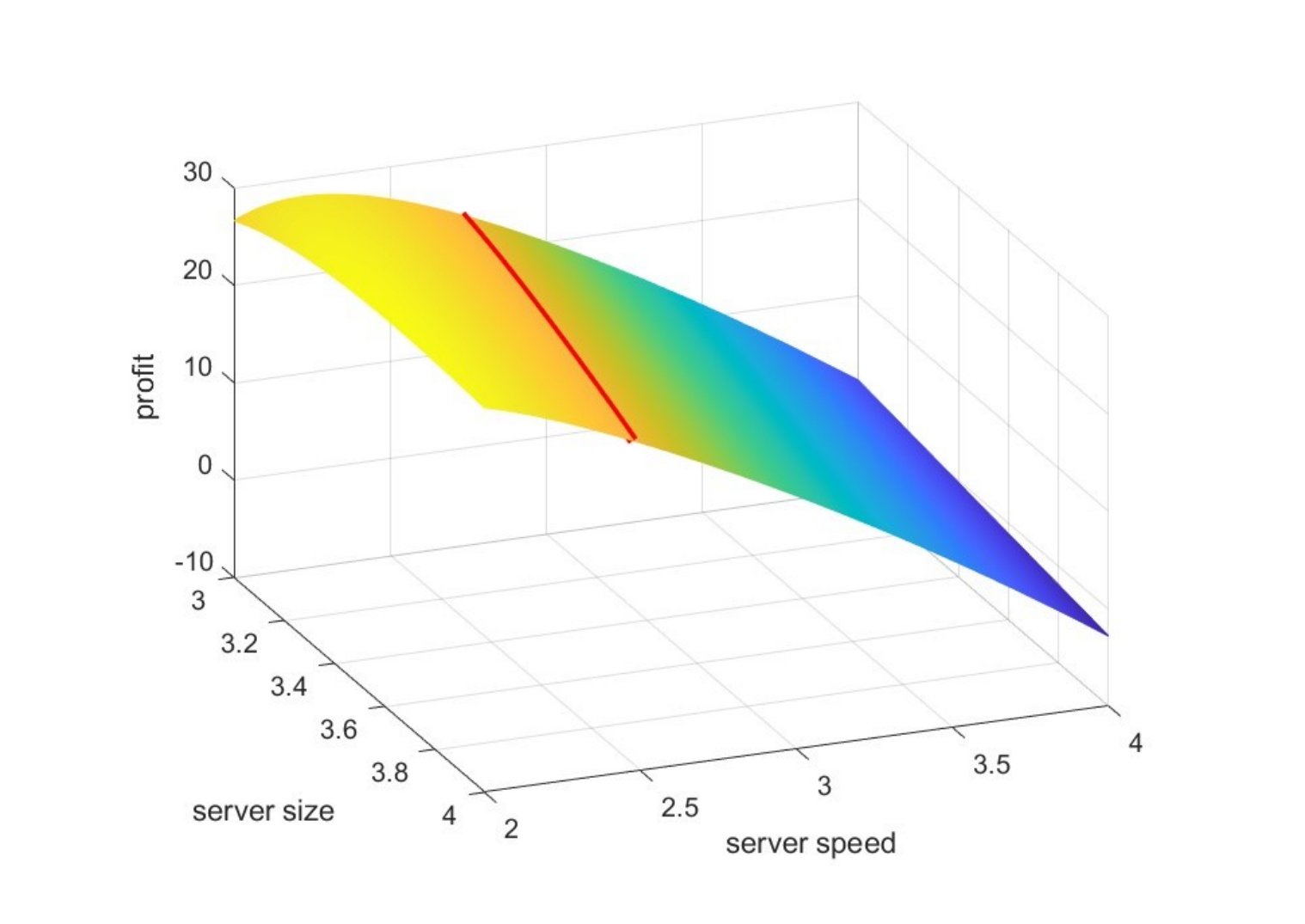} }
  \caption{The mesh of profit $G$ versus $s$ and $m$.}
   \label{Fig:Fig3}
\end{figure}

By setting $\lambda  = 4$ hundred service requests per hour, ${a} = 15$  dollars per one hundred billion instructions (Note: The monetary unit “dollars” in this paper may not be identical but should be linearly proportional to the real  US dollars.), $\beta  = 3$ dollars per one hundred billion instructions, $\delta  = 1$ and ${P^*} = 4$, and then traversing the size and speed of the servers in $[3,4]$ and $[2,4]$, we draw the three-dimensional surfaces of \eqref{Eqs:Eqs20} and \eqref{Eqs:Eqs22}, which are shown in Figure.\ref{Fig:Fig2} and Figure.\ref{Fig:Fig3} respectively. As can be seen from Figure.\ref{Fig:Fig2}, when the size and speed of the servers decrease simultaneously, customers have to take a longer waiting time for their requests to be served, and vice versa. While for Figure.\ref{Fig:Fig3}, when the size and speed of the servers are big, the service ability of the servers are so strong that it will far exceed the demand of customers, which will result in the larger amount of costs of cloud service providers than their revenues. As the size and speed of the servers decrease, less waste of the service ability will bring higher profits to cloud service providers. However, when the size and speed of the servers further decrease, the profit of cloud service providers will decrease instead, for the reason that the service ability is too weak to serve enough requests before the deadline, such that parts of the customers will enjoy the services for free.

\begin{figure}[!t]\centering
  \centerline{\includegraphics[width=0.53\textwidth]{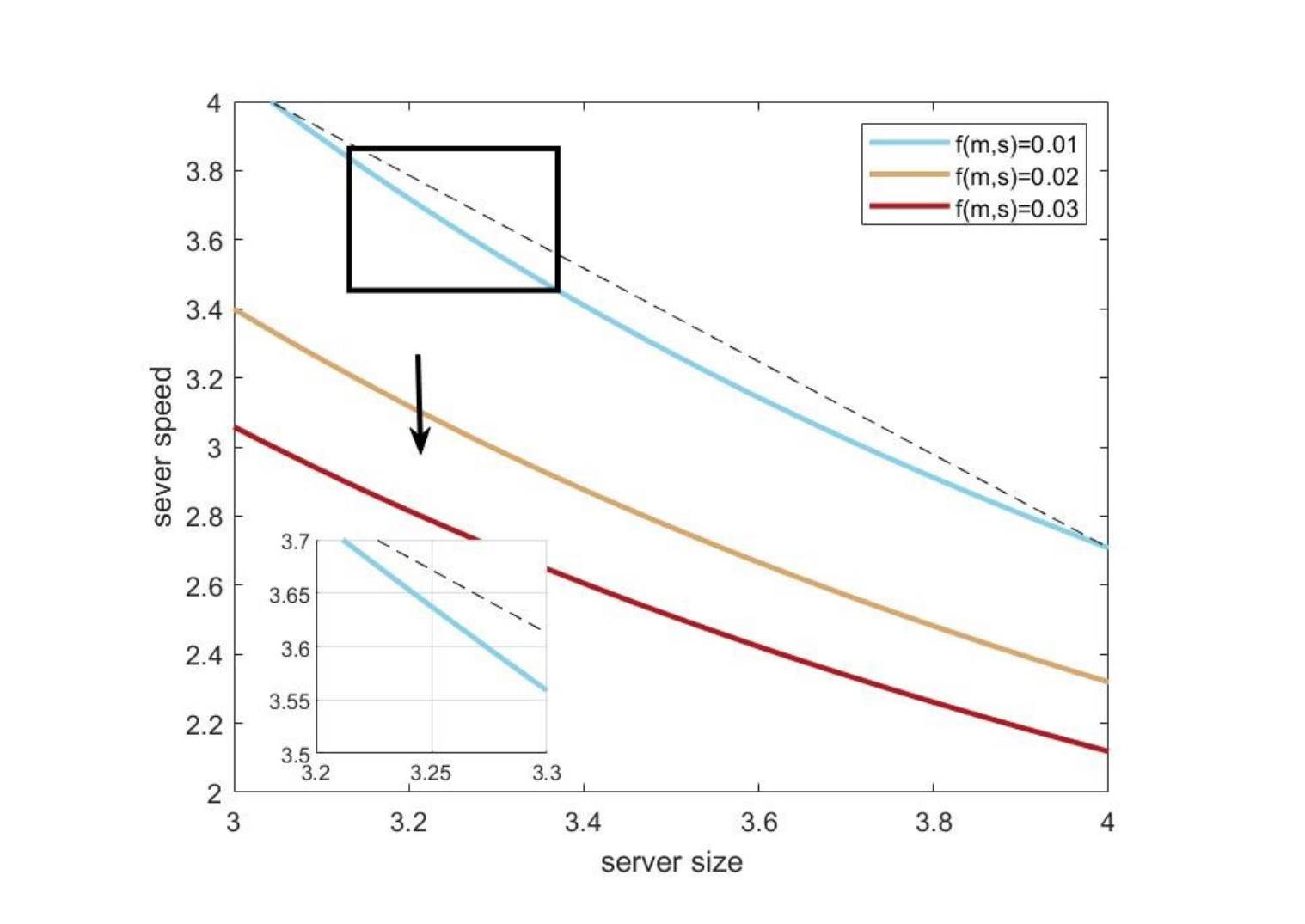} }
  \caption{Deadline $D$ versus $s$ and $m$.}
   \label{Fig:Fig4}
\end{figure}

Consider the definite value of the waiting time of customers in Figure.\ref{Fig:Fig2}, i.e., $f\left( {m,s} \right) = {c_1}$, where ${c_1}$ is a constant, we can find the implicit function relationship between the size and speed of the servers in \eqref{Eqs:Eqs20}, which can be plotted as the red curve in Figure.\ref{Fig:Fig2}. By specifying $c_1 \in \left[ {0.01, 0.02, 0.03} \right]$, and mapping the implicit function relationship into $\left( {m,s} \right)$ plane, three curves can be drawn in Figure.\ref{Fig:Fig4}. On this basis, when we set deadline as $D = 0.01$, only if the size and speed of the servers are selected in the region above the curve bottom in Figure.\ref{Fig:Fig4} can the customer requests be served before the deadline. In other words, such region can be named as the $feasible \ region$ with the waiting time of customers being taken into consideration.

\begin{figure}[!t]\centering
  \centerline{\includegraphics[width=0.53\textwidth]{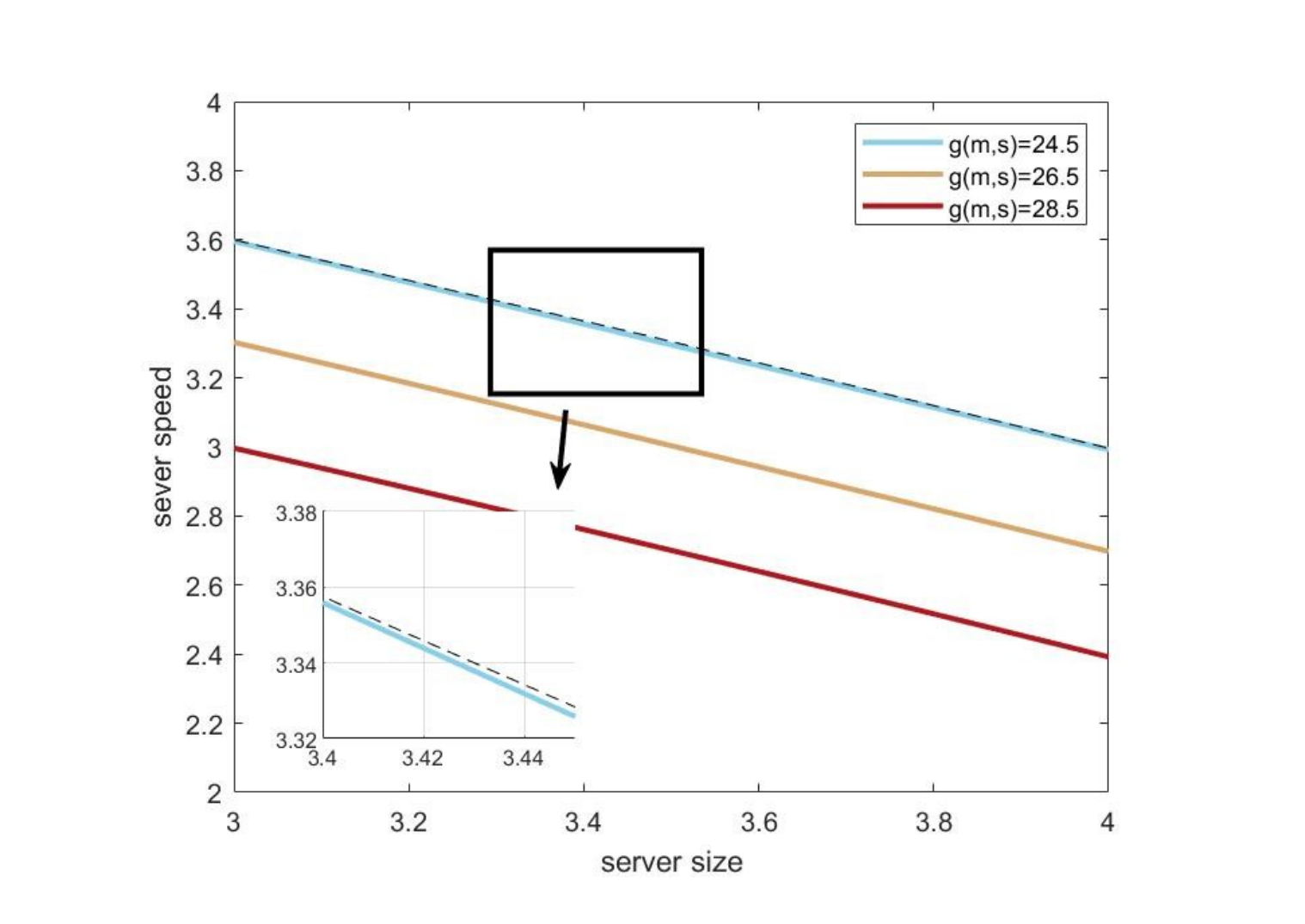} }
  \caption{Profit $G$ versus $s$ and $m$.}
   \label{Fig:Fig5}
\end{figure}

Similarly, consider the definite value of the profit of cloud service providers in Figure.\ref{Fig:Fig3}, i.e., $g\left( {m,s} \right) = {c_2}$, where ${c_2}$ is a constant, we can find the implicit function relationship between the size and speed of the servers in \eqref{Eqs:Eqs22}, which can be plotted as the red curve in Figure.\ref{Fig:Fig3}. By specifying $c_2 \in \left[ {24.5,26.5,28.5} \right]$, and mapping the implicit function relationship into $\left( {m,s} \right)$ plane, three curves can be drawn in Figure.\ref{Fig:Fig5}. On this basis, when we set the acceptable lowest profit as $G = 24.5$, only if the size and speed of the servers are selected in the region below the curve top in Figure.\ref{Fig:Fig5} can the cloud service providers gain more profits. In other words, such region can be named as the $feasible \ region$ with the profit of cloud service providers being taken into consideration.

\begin{figure}[!t]\centering
  \centerline{\includegraphics[width=0.53\textwidth]{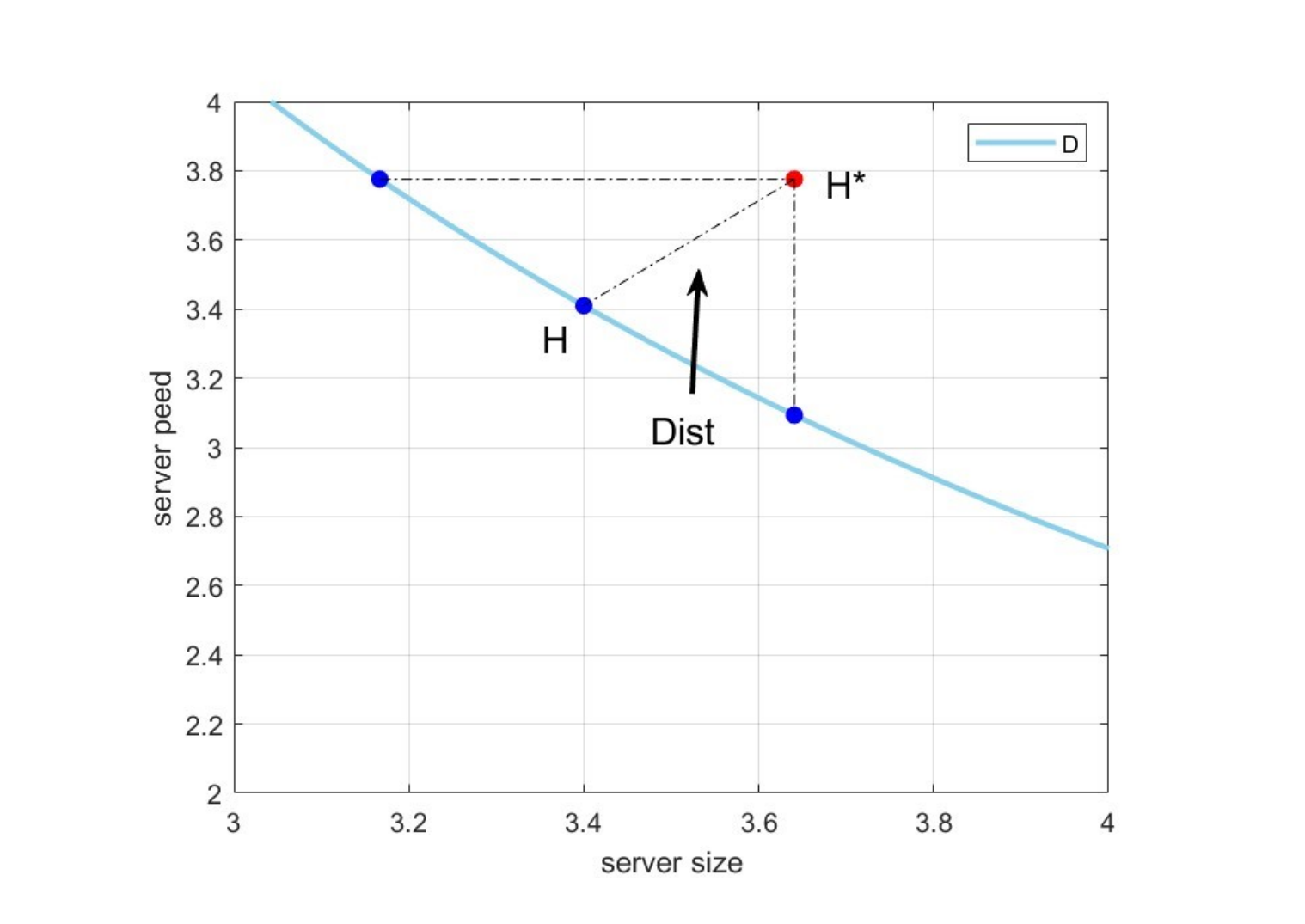} }
  \caption{The $working \ point$ $H^*$ versus to $D$.}
   \label{Fig:czt}
\end{figure}

Based on the above discussion, the impacts of the perturbation factors on the profit of cloud service providers and the waiting time of customers will be described. Now we draw a certain point in the $feasible \ region$ of Figure.\ref{Fig:Fig4}, which is shown as the red point in Figure.\ref{Fig:czt}. In this case, we called the red point as the $working \ point$, which means the cloud platform is operating with the size and speed of servers being set as the horizontal and vertical coordinates of this point respectively. When the perturbation happens, the size and speed of servers will deviate from the $working \ point$. If the deviation degree is small, the new working point will still locate in the $feasible \ region$. However, if the deviation degree tends to increase, the waiting time of customers may be greater than the deadline, which is unacceptable for the demand of customers.

\begin{figure}[!t]\centering
  \centerline{\includegraphics[width=0.53\textwidth]{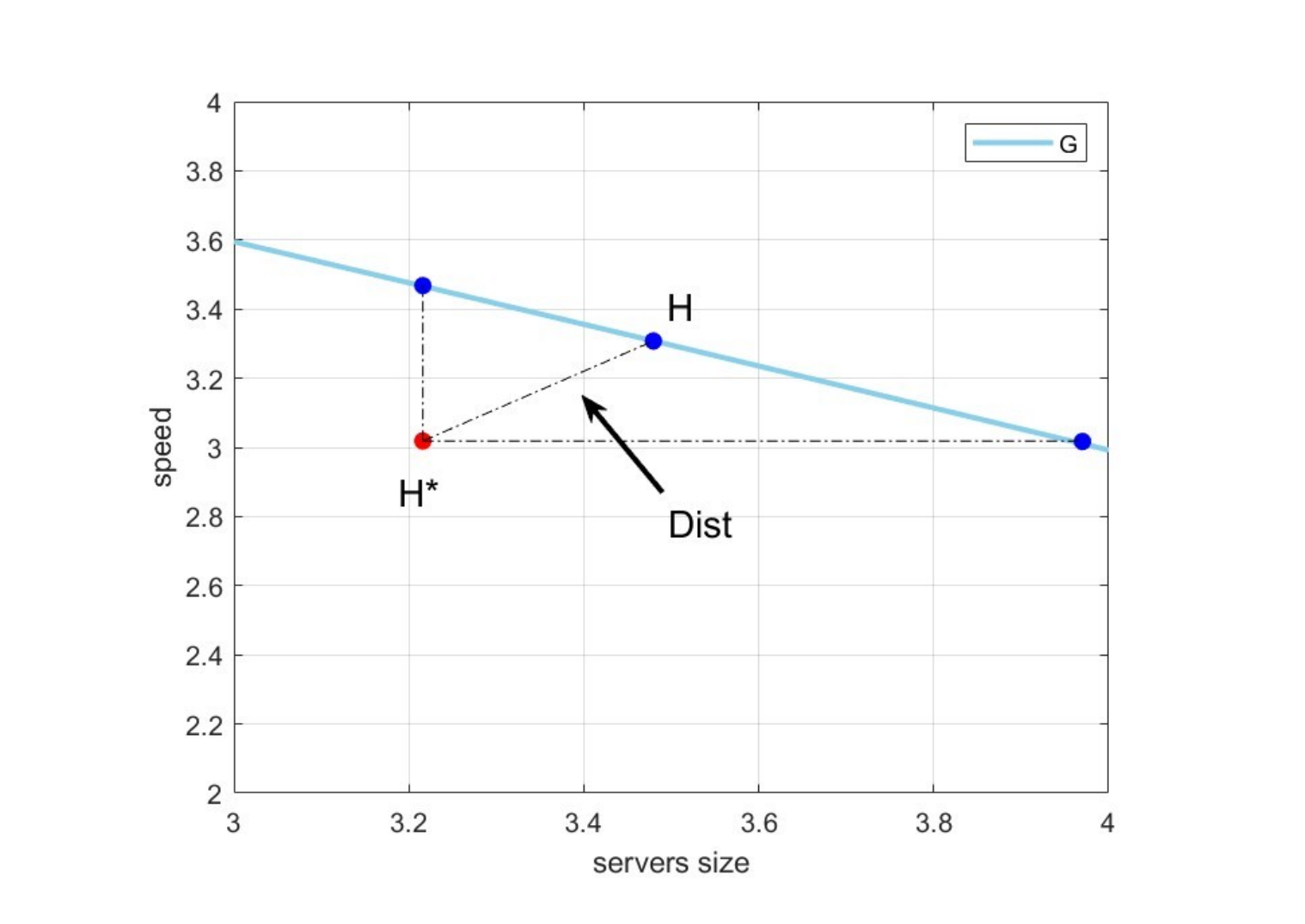} }
  \caption{The $working \ point$ $H^*$ versus to $G$.}
   \label{Fig:czg}
\end{figure}

Moreover, we draw a certain point in the $feasible \ region$ of Figure.\ref{Fig:Fig5}, which is shown as the red point in Figure.\ref{Fig:czg}. Similar to the point defined in Figure.\ref{Fig:czt}, we also called the red point as the $working \ point$. When the perturbation happens in this case, the size and speed of servers will deviate from the $working \ point$. If the deviation degree is small, the new working point will still locate in the $feasible \ region$. However, if the deviation degree tends to increase, the profits of cloud service providers may be less than their acceptable lowest value, which is unacceptable for the demand of cloud service providers.

In order to overcome the problems for customers and cloud service providers, by introducing the idea of robustness, we try to promote the performance of cloud platform as much as possible when it is operating in the worst case. On this basis, any other cases better than the current one will lead to better performance of cloud platform. Specifically, consider Figure.\ref{Fig:czt} and Figure.\ref{Fig:czg} respectively, we can denote the $working \ point$ as $H^{*}$, and any point on the curves as $H$, then the distance between $H^{*}$ and $H$ can be defined as follow.

\begin{equation}\label{Eqs:Dis}
    Dist = {\left\| {{H^*} - H} \right\|_2}
\end{equation}

In addition, given any $working \ point$ in the $feasible \ region$ in Figure.\ref{Fig:czt} and Figure.\ref{Fig:czg}, we can find the shortest distance between such point and the point on the curves. In this case, due to the deviation in the size and speed of servers, we think the cloud platform has the highest probability to operate in the unacceptable region when the perturbation happens. Therefore, we need to find the best $working \ point$ in the $feasible \ region$, such that the shortest distance between such point and the point on the curves can be enlarged as much as possible.

\section{Robustness analysis}\label{sec: Robustness analysis}

In this section, we focus on designing an adequate algorithm for determining the shortest distance between the $working \ point$ and the points on the boundaries, namely, the curves in Figure.\ref{Fig:czt} and Figure.\ref{Fig:czg} respectively. Then, by applying such algorithm to any $working \ point$ in the $feasible \ region$, the horizontal and vertical coordinates of the point with the longest "shortest distance" will be chosen as the optimal value for the configuration of server size and server speed in cloud platform.

\begin{figure}[!t]\centering
  \centerline{\includegraphics[width=0.53\textwidth]{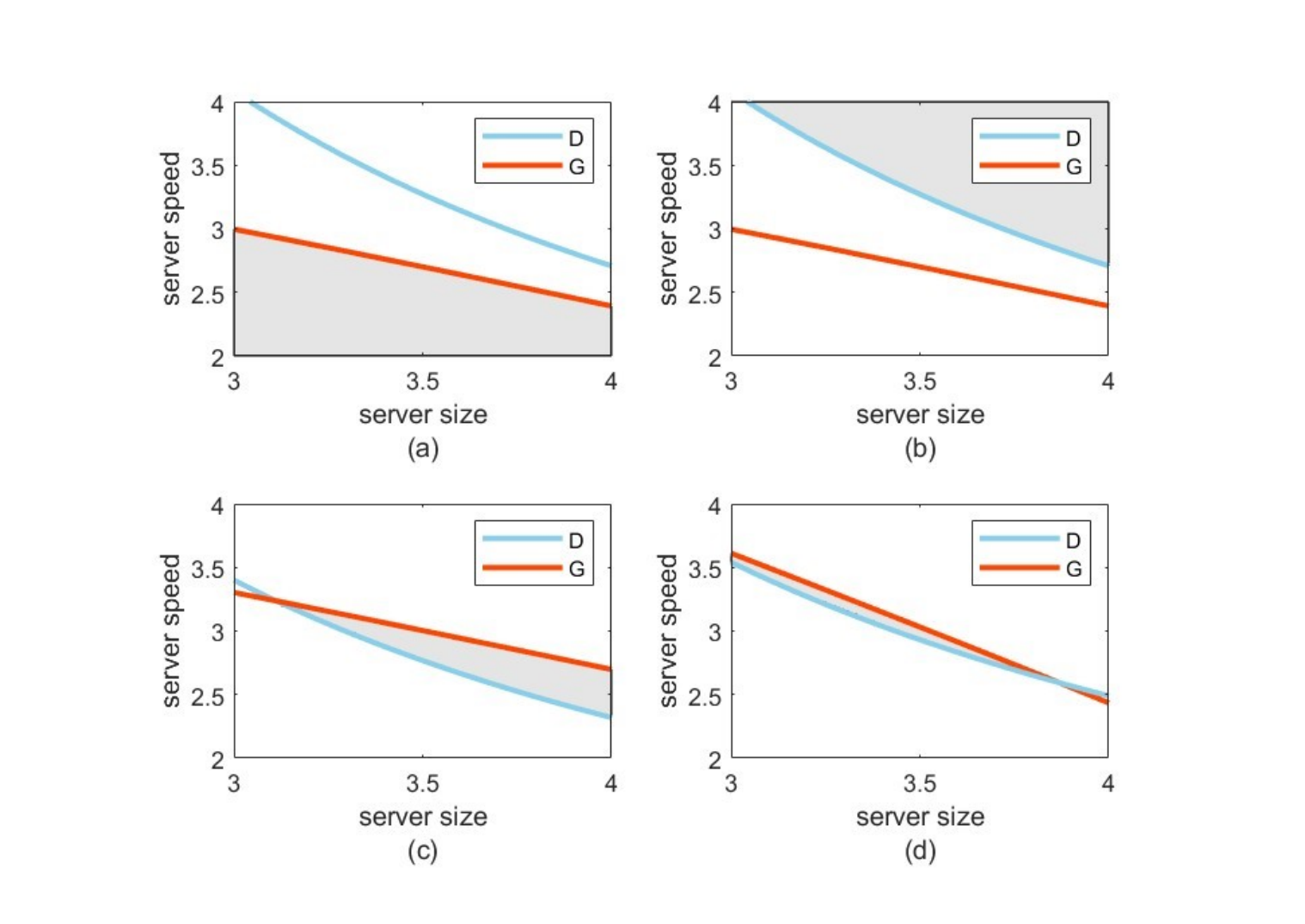} }
  \caption{The $feasible \ region$ related to  $G$ and $D$.}
   \label{Fig:bj}
\end{figure}

Based on the above discussion, the $feasible \ region$ in Figure.\ref{Fig:czt} and Figure.\ref{Fig:czg} can be combined into one graph, which is shown in Figure.\ref{Fig:bj}. Regarding the fact that the curves in any sub-graph have intersected with each other or not, those sub-graphs in Figure.\ref{Fig:bj} can be separated into two cases. Intuitively, Figure.\ref{Fig:bj} (a) and (b) show the fact that the curves have not intersected with each other, while Figure.\ref{Fig:bj} (c) and (d) show the opposite result.

Notice that, the blue curve and yellow curve in Figure.\ref{Fig:bj} implies the boundaries for the profit of cloud service providers and the waiting time of customers respectively. Then, recalling that the $feasible \ region$ for the profit of cloud service providers is the region below the blue curve, and the $feasible \ region$ for the waiting time of customers is the region above the yellow curve, we can easily find that one of these regions has not overlapping with another in Figure.\ref{Fig:bj} (a) and (b), while such overlaps are existed in Figure.\ref{Fig:bj} (c) and (d). For the former case, the $working \ point$ can not be uniquely determined in the $feasible \ region$ both for the profit of cloud service providers and the waiting time of customers simultaneously. Therefore, we have to search the $working \ point$ in the $feasible \ region$ for the profit of cloud service providers in Figure.\ref{Fig:bj} (a), and then search the $working \ point$ in the $feasible \ region$ for the waiting time of customers in Figure.\ref{Fig:bj} (b) respectively. For the latter case, the shaded parts in Figure.\ref{Fig:bj} (c) and (d) show the overlapping region, which permit us to find the $working \ point$ suitable for the $feasible \ region$ both for the profit of cloud service providers and the waiting time of customers simultaneously.

\subsection{Robustness metric}

Now, we devote to calculate the shortest distance from the $working \ point$ to the boundaries with respect to the profit of cloud service providers and the waiting time of customers respectively. Since the principle for searching the optimal $working \ point$ is maximizing the previous mentioned shortest distance, which is similar to the idea of robustness described in the control field, it is reasonable to rename such distance as the robustness radius\cite{ali2004measuring}, which can be represented as follow.

\begin{equation}\label{Eqs:Robust}
    {r^*} = \max \left\{ {min\left\{ {Dist} \right\}} \right\} = \mathop {\max }\limits_{{H^*}} \mathop {min}\limits_H {\left\| {{H^*} - H} \right\|_2}
\end{equation}

\begin{figure}[!t]\centering
  \centerline{\includegraphics[width=0.53\textwidth]{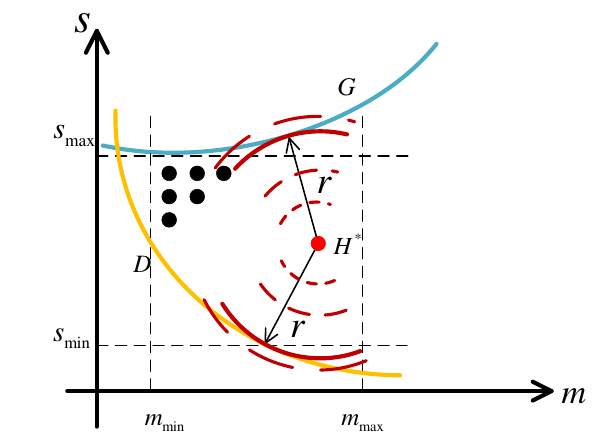} }
  \caption{Shortest robustness radius $r$ from $H^*$ to $G$ or $D$.}
   \label{Fig:js}
\end{figure}

Given a specific $working \ point \ H^*$  in $feasible \ region$, which is shown in Figure.\ref{Fig:js}, it is easily to find that the mathematical expression of the shortest distance from such point to the boundary can hardly be deduced, for the reason that the formulas for the profit of cloud service providers and the waiting time of customers is very complicated. Therefore, we assume that a circle whose center is $H^*$ and radius $r$ is the shortest distance from $H^*$ to the boundary will be tangent to such boundary. In this case, we propose an adaptive adjustment scheme for the radius $r$, such that the circle centered on $H^*$ can be tangent to the boundary. The pseudo code for such scheme is shown in Algorithm.\ref{alg1}.

Considering a circle centered on $H^*$, whose horizontal and vertical coordinates are $m_z$ and $s_z$ respectively, such circle can be represented in polar coordinates. As can be seen from Line 5 and 6 in Algorithm.\ref{alg1}, $\boldsymbol{\theta}$ is the polar angle in vector form, the elements in the vector take the value between 0 to $2\pi$ uniformly. By means of the equations, $\boldsymbol{x}$ and $\boldsymbol{y}$ can be calculated as the coordinates of the points on the circle in vector form. Then, let $l(x,y)$ to be an unified form of $g(x,y)$ and $f(x,y)$, Line 7 can be adopted to determine whether the constructed circle has intersected with the boundaries or not. If it intersects, the radius of the circle should be decreased. In this case, we define two indicator functions to distinguish $g(x,y)$ from $f(x,y)$ in the unified form, which are shown in \eqref{Eqs:zhi1} and \eqref{Eqs:zhi2}. Line 9 demonstrates that whether the constructed circle is tangent to the boundaries or not. If there has no tangency, the radius of the circle should be increased. Otherwise, we will try to find the tangent point on the circle and the shortest radius from $working \ point$ to the boundaries, which can be implemented by Line 12 to 20. Notice that, $h(x,y)$ denotes the tangent equation of $l(x,y)$ due to the implicit function property with respect to the relationship between $m$ and $s$ in \eqref{Eqs:Eqs20} and \eqref{Eqs:Eqs22}, then $-(x_i-m_z)/(y_i-s_z)$ denotes the tangent equation of the constructed circle whose horizontal and vertical coordinates are $x_i$ and $y_i$ respectively.

\begin{algorithm}[t]
    \caption{Shortest radius $r$ from $working \ point$ to the boundary}
    \label{alg1}
    \LinesNumbered
    \DontPrintSemicolon
    \KwIn{ ${m_z}$, ${s_z}$, $r_{max}$, $\boldsymbol{\theta}$, $\upsilon_1$, $\upsilon_2$, $\xi_1$, $\xi_2$, $\mathbbm{1}_1$, $\mathbbm{1}_2$, $c$}
    \KwOut{Shortest radius $r$ from $working \ point$ to the boundary}
    \Begin{
        $\xi_1,\xi_2  \leftarrow 10^{-4}$;
        $\upsilon_1,\upsilon_2 \leftarrow 10^{-5}$; \\
        $r_1 \leftarrow 10^{-4}$; $r_{max}  \leftarrow 10^{4}$ ;\;
        \While{$r_1<r_{max}$}{
            $\boldsymbol{x} = {m_z} + r_1 \cdot \cos \boldsymbol{\theta} $;\;
            $\boldsymbol{y} = {s_z} + r_1 \cdot \sin \boldsymbol{\theta} $;\;
              
            \uIf{$\sum\nolimits_{i = 1}^n {\left| {\boldsymbol{l}\left( {\boldsymbol{x},\boldsymbol{y}} \right) - c} \right|}  \ne \left| {\sum\nolimits_{i = 1}^n {\left[ {\boldsymbol{l}\left( {\boldsymbol{x},\boldsymbol{y}} \right) - c} \right]} } \right|$}
            {
                $r_1=r_1-\mathbbm{1}_{1} \cdot v_1-\mathbbm{1}_{2} \cdot v_2$;\;
            }
            \uElseIf{$\prod_{i=1}^n{\left\{ \boldsymbol{l}\left( \boldsymbol{x},\boldsymbol{y} \right) -c \right\} \ne 0}$}
            {
                $r_1=r_1+\mathbbm{1}_{1} \cdot \xi _1+\mathbbm{1}_{2} \cdot \xi _2$;\;
            }
              \Else{
                 \For{${\theta_i=0}$ to ${2\pi}$}{
                 $x_i = {m_z} + r_1 \cdot \cos \theta_i $;\;
                 $y_i = {s_z} + r_1 \cdot \sin \theta_i $;\;
                 $h({x},{y})=-\frac{l_{{x}}^{'}}{l_{{y}}^{'}}$;\;

                 \If{  $h(x_i,y_i)==-\frac{x_i-m_z}{y_i-s_z}$ $\land$ $l\left( {x_i,y_i} \right) == c$}
             {$r \leftarrow r_1$;\\
             break; }
                 
                 }
    }
}
\Return{$r$}
}
\end{algorithm}

\begin{equation}\label{Eqs:zhi1}
{{\mathbbm{1}_1}} = \left\{ {\begin{array}{*{20}{c}}
  {0,}&{if \quad l\left(x,y\right)=g\left(x,y\right)} \\ 
  {1,}&{else}
\end{array}} \right.
\end{equation}

\begin{equation}\label{Eqs:zhi2}
{{\mathbbm{1}_2}} = \left\{ {\begin{array}{*{20}{c}}
  {0,}&{if \quad l\left(x,y\right)=f\left(x,y\right)} \\ 
  {1,}&{else}
\end{array}} \right.
\end{equation}

\subsection{Situation A}


In the previous subsection, we design an algorithm to obtain the shortest distance between a specific $working \ point$ and the points on the given boundaries. On this basis, we aim to find a $working \ point$ with robustness radius from such point to the boundaries in the $feasible \ region$, namely, by means of such method, the longest "shortest distance" defined in \eqref{Eqs:Robust} can be obtained. Now, we first consider the case described in Figure.\ref{Fig:bj} (a), and then, the analysis procedure in such case can be extended into other cases described in Figure.\ref{Fig:bj} (b), (c) and (d) respectively.

\subsubsection{Optimal size and speed}

As can be seen from Figure.\ref{Fig:bj} (a), the dashed region below the boundary with respect to the profit of cloud service providers is considered as $feasible \ region$. In this case, compared to the waiting time of customers, we pay attention to seek a profit oriented greedy strategy, such that cloud service providers can earn as much profits as they can. \\
\indent
 To achieve the target chased by cloud service providers, which is equivalent to find the optimal $working \ point$ with robustness radius in $feasible \ region$, we introduce Dung Beetle Optimizer (DBO) algorithm \cite{xue2022dung} to find such optimal $working \ point$, in which the size and speed of the servers, i.e., the horizontal and vertical coordinates of $working \ point$ are selected as the optimization variables of DBO algorithm. The pseudo code for such algorithm is shown in Algorithm.\ref{alg2}.

\begin{algorithm}[t]
    \caption{Robust scheduling method based on DBO algorithm}
    \label{alg2}
    \LinesNumbered
    \DontPrintSemicolon

    \KwIn{The maximum iterations $T_{max}$, the size of the  population $N$}
         
\KwOut{Optimal position $x^z$and its robustness radius $r$}
               Initialize the particle's population $i\leftarrow 1,2,...,N$; $x^z \leftarrow$ two dimensions $m_z$, $s_z$;  Define its relevant parameters $l{b^{'}}$, $u{b^{'}}$, $c_1$, $c_2$, $S$, $g$;\;
         \While{$t<t_{max}$}{
           \For{ $i \leftarrow 1$ to $N$ }{
             \If{i is ball-rolling dung beetle}{
               $\delta \leftarrow rand(1)$; \;  
             
                \uIf{$\delta  < 0.9$}{
                  $\alpha  \leftarrow $through the  $\alpha$ selection strategy; \;
                $x_i^{t + 1} = x_i^t + \alpha  \cdot k \cdot x_i^{t - 1} + b \cdot \left| {x_i^t - x_{worst}^t} \right|$;\;
                Obtain the $r_{i}$ corresponding to $x_{i}^{t + 1}$ by Algorithm 1;\;
                }
                \Else{
                $\theta  \leftarrow $through the  $\theta$ selection strategy; \;
               $x_i^{t + 1} = x_i^t + \tan \theta \left| {x_i^t - x_i^{t - 1}} \right|$;\;
               Obtain the $r_{i}$ corresponding to $x_{i}^{t + 1}$ by Algorithm 1;\;
                }
            }
       
             \If{i is brood ball}{
             Restrict the region with a boundary selection strategy based on the current optimal location $x_{gbest}^{'}$, update the location;\;
             Obtain the $r_{i}$ corresponding to $x_{i}^{t + 1}$ by Algorithm 1;
             }
             \If{i is small dung beetle}
             {$x_i^{t + 1} = x_i^{t} + {c_1}  \left( {x_i^{t} - l{b^{'}}} \right) + {c_2}  \left({x_i^{t} - u{b^{'}}}\right)$;\;
                Obtain the $r_{i}$ corresponding to $x_{i}^{t + 1}$ by Algorithm 1;\;
             }
             \If{i is thief}{
             $x_i^{t + 1} = x_{lbest}^{t}t + S \cdot g \cdot \left( {\left| {x_i^{t} - x_{gbest}^{t}} \right| + \left| {x_i^{t} - x_{lbest}^{t}} \right|} \right)$;\;
             Obtain the $r_{i}$ corresponding to $x_{i}^{t + 1}$ by Algorithm 1;
             }
         }
    \If{the newly generated $r$ is better than before}{
                  $r \leftarrow r_{i}$;
                  $x^z \leftarrow x_i^{t + 1}$;
            }
            $t \leftarrow t+1$; \;  
              }
      
\end{algorithm}

\begin{figure}[!t]\centering
  \centerline{\includegraphics[width=0.53\textwidth]{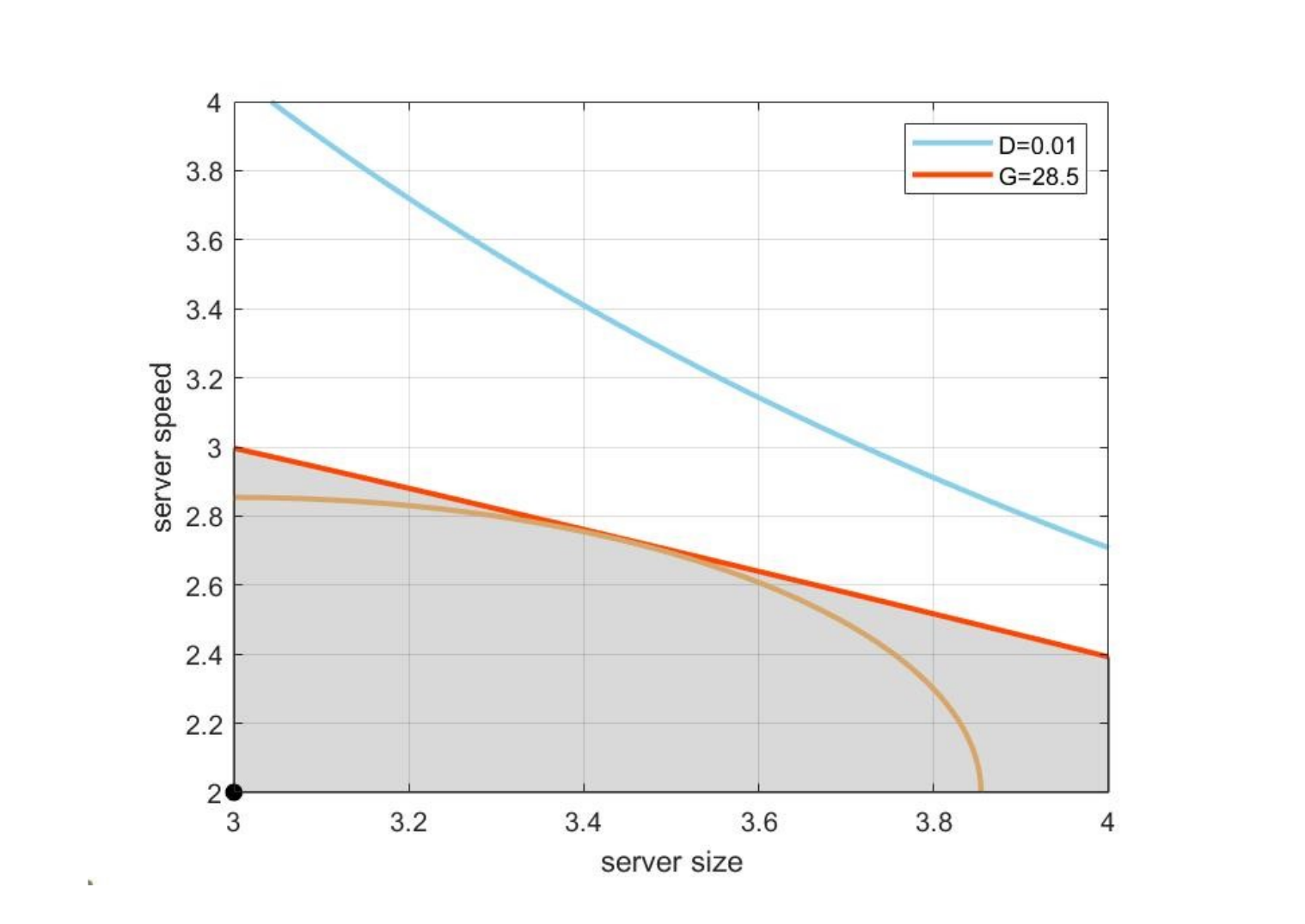} }
  \caption{The $feasible \ region$ formed by $G$.}
   \label{Fig:qk1}
\end{figure}

By setting the acceptable lowest profit as $G = 28.5$ to represent the boundary for the profit of cloud service providers, and adopting DBO algorithm, the corresponding result can be shown in Figure.\ref{Fig:qk1}. Apparently, the optimal $working \  point$ can be found in the bottom-left corner of the figure with minimum size and speed of the servers in the $feasible \ region$. Namely, when we choose the size and speed of servers as $m_z = 3$ and $s_z = 2$ respectively, the robustness radius can be calculated as $r = 0.852$ to represent the maximized shortest distance from the $working \  point$ to the boundary. When refers to a more intuitive explanation for the position of the optimal $working \  point$, the smaller size and speed of servers configured by cloud platform, the less costs will be spent by cloud service providers, however, the payments of customers always remain the same due to the flat pricing strategy, which will bring more profits to cloud service providers.

\indent

\subsubsection{Performance analysis}

\begin{figure}[!t]\centering
  \centerline{\includegraphics[width=0.53\textwidth]{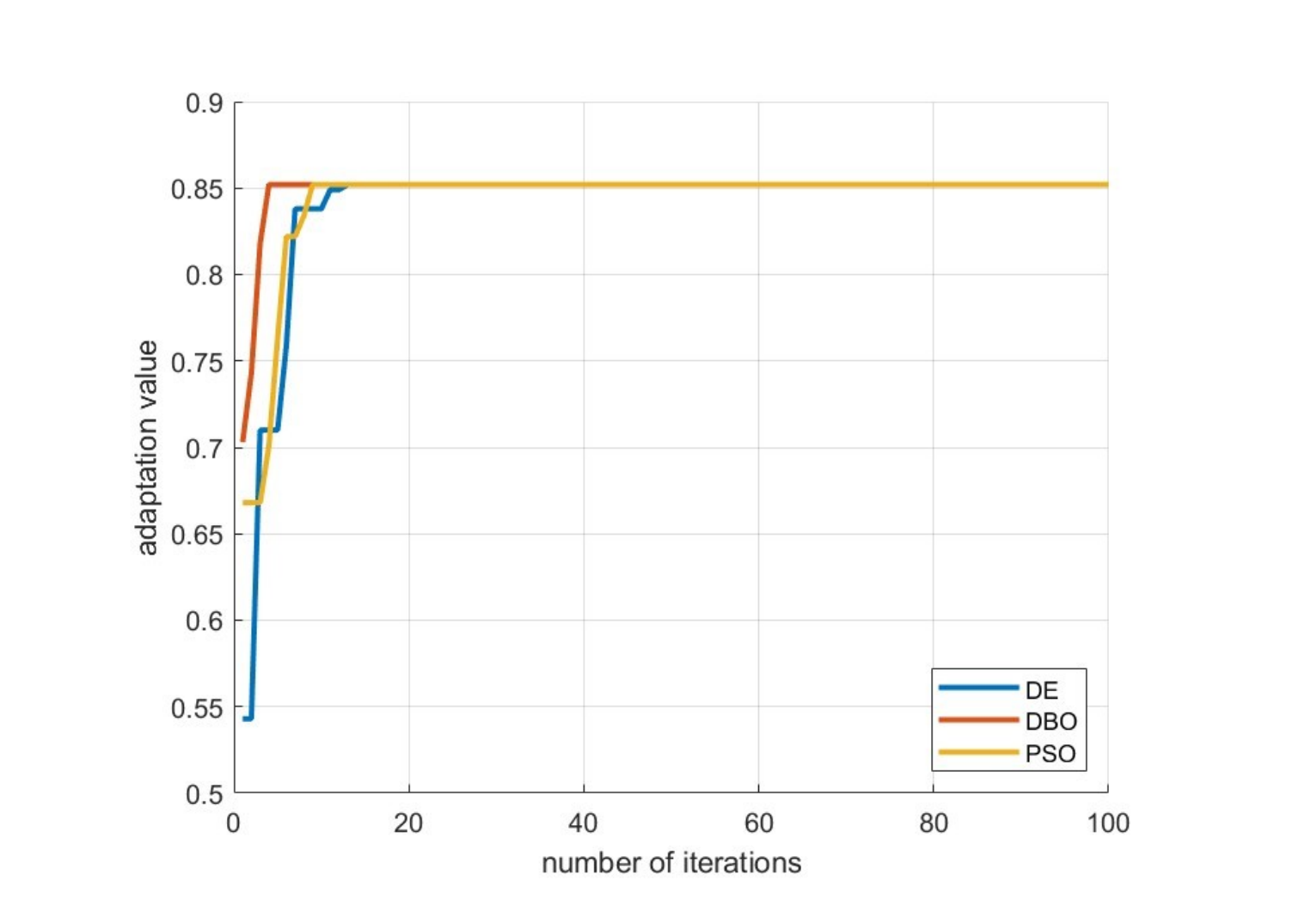} }
  \caption{Shortest robustness radius $r$ for situation A.}
   \label{Fig:sf1}
\end{figure}

In order to go verify the superiority of DBO algorithm, Differential Evolution (DE) algorithm and Partical Swarm Optimization (PSO) algorithm are adopted to obtain robustness radius as well. As can be seen from Figure.\ref{Fig:sf1}, the shortest distance between $working \ point$ and boundary has been chosen as the fitness function of three algorithms. By simulating such algorithms, all of these algorithms can attain the same robustness radius. However, the shortest distance between $working \ point$ and boundary converges to the optimal value by DBO algorithm with the least number of iterations.

\begin{figure}
  \centering
   \centerline{\includegraphics[width=0.53\textwidth]{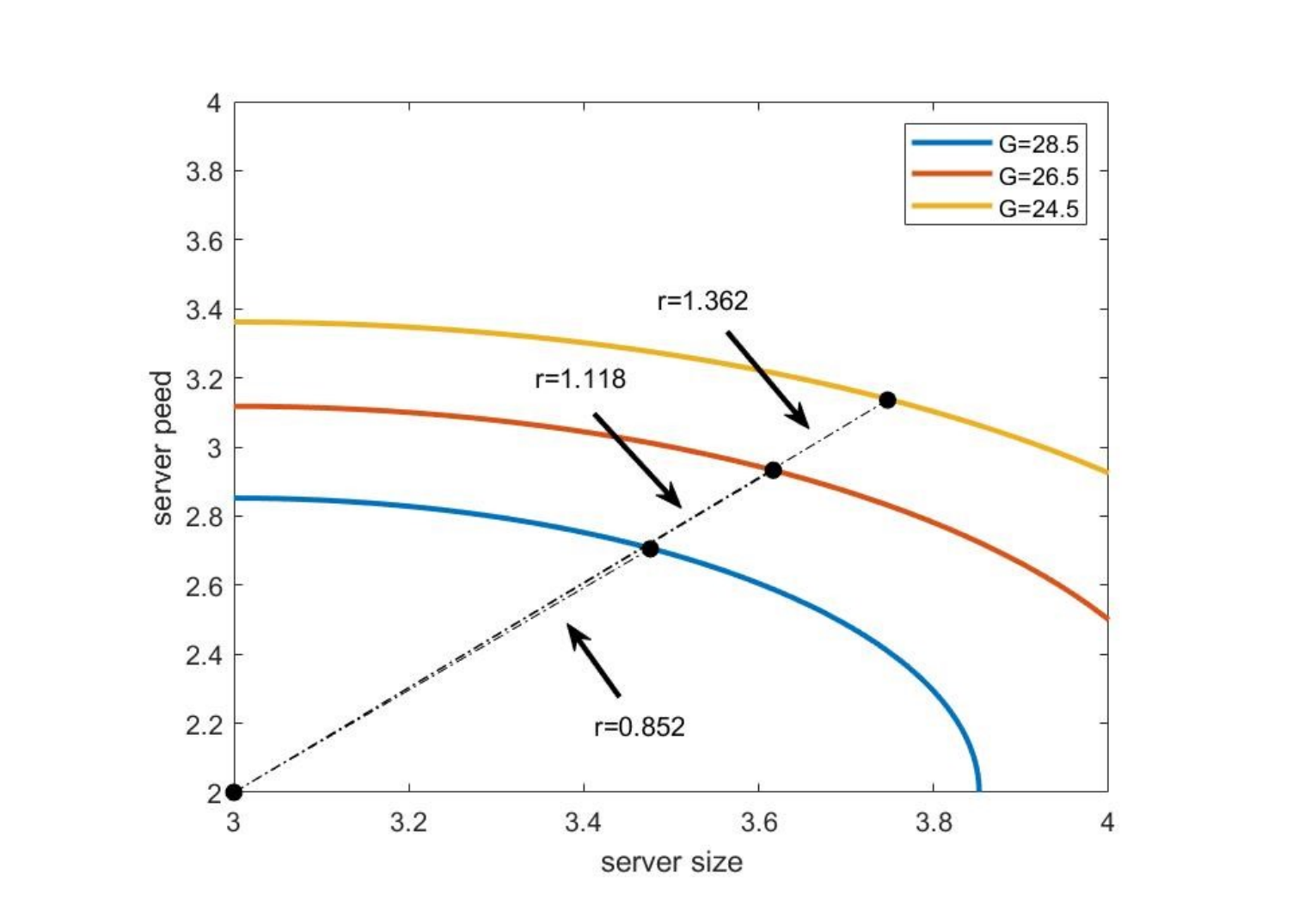} }
  \caption{Shortest robustness radius $r$ versus $G$.}
  \label{Fig:dg32}
\end{figure}

Furthermore, we perform the robustness analysis for the case presented in Figure.\ref{Fig:bj} (a) with the acceptable lowest profit being setting to $G=24.6$, $G=26.5$ and $G=28.5$, respectively. By simulating the proposed algorithm, the optimal $working \ point$ can be obtained at the same position for three cases, where the size and speed of servers are $m_z = 3$ and $s_z = 2$. As can be seen from Figure.\ref{Fig:dg32}, the robustness radius for $G=24.5$, $G=26.5$ and $G=28.5$ cases are $1.362$, $1.118$ and $0.852$ respectively, for the reason that the higher the acceptable lowest profit is, the larger possibility the cloud environment be affected by the perturbations in the size and speed of servers.

\subsection{Situation B}

Now, compared to the profit of cloud service providers, we turn to seek a waiting time oriented greedy strategy, such that the waiting time that customers spend can be decreased as much as possible. As can be seen from Figure.\ref{Fig:bj} (b), the dashed region above the boundary with respect to the waiting time of customers is considered as $feasible \ region$.

\subsubsection{Optimal size and speed}

\begin{figure}[!t]\centering
  \centerline{\includegraphics[width=0.53\textwidth]{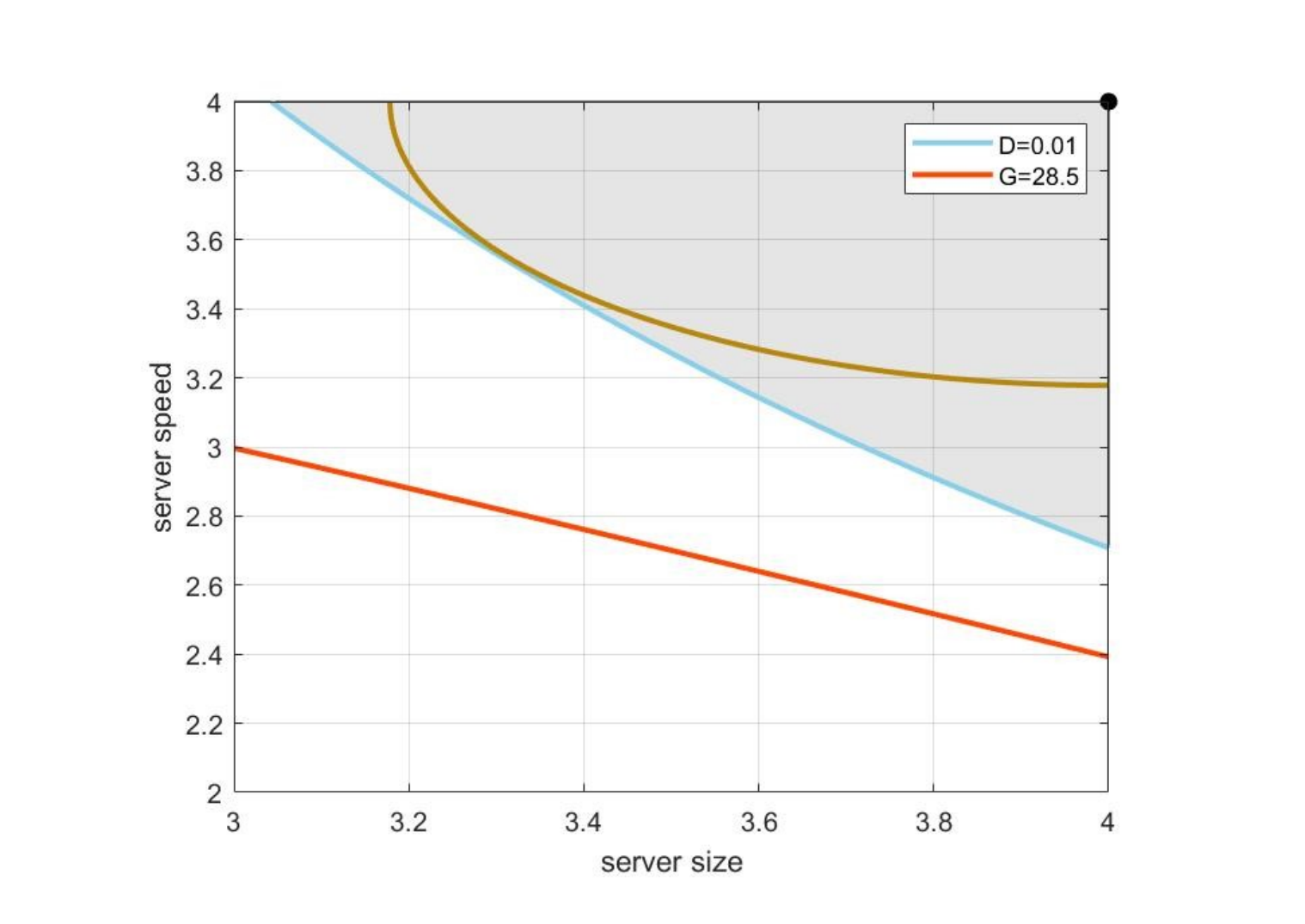} }
  \caption{The $feasible \ region$ formed by $D$.}
   \label{Fig:qk3}
\end{figure}

Similar to the method used in the previous subsection, to achieve the target chased by customers, we also adopt DBO algorithm to find the optimal $working \ point$ with robustness radius in $feasible \ region$. Then, by setting the deadline as $D = 0.01$ to represent the boundary for the waiting time of customers, the result simulated by DBO algorithm can be shown in Figure.\ref{Fig:qk3}. Apparently, the optimal $working \  point$ can be found in the upper right corner of the figure with maximum size and speed of the servers in the $feasible \ region$. Namely, when we choose the size and speed of servers as $m_z = 4$ and $s_z = 4$ respectively, the robustness radius can be calculated as $r = 0.777$ to represent the maximized shortest distance from the $working \  point$ to the boundary. When refers to a more intuitive explanation for the position of the optimal $working \  point$, the bigger size and speed of servers configured by cloud platform, the higher efficiency for the service requests can be handled, which will decrease the waiting time of customers effectively. 

\subsubsection{Performance analysis}

\begin{figure}[!t]\centering
  \centerline{\includegraphics[width=0.53\textwidth]{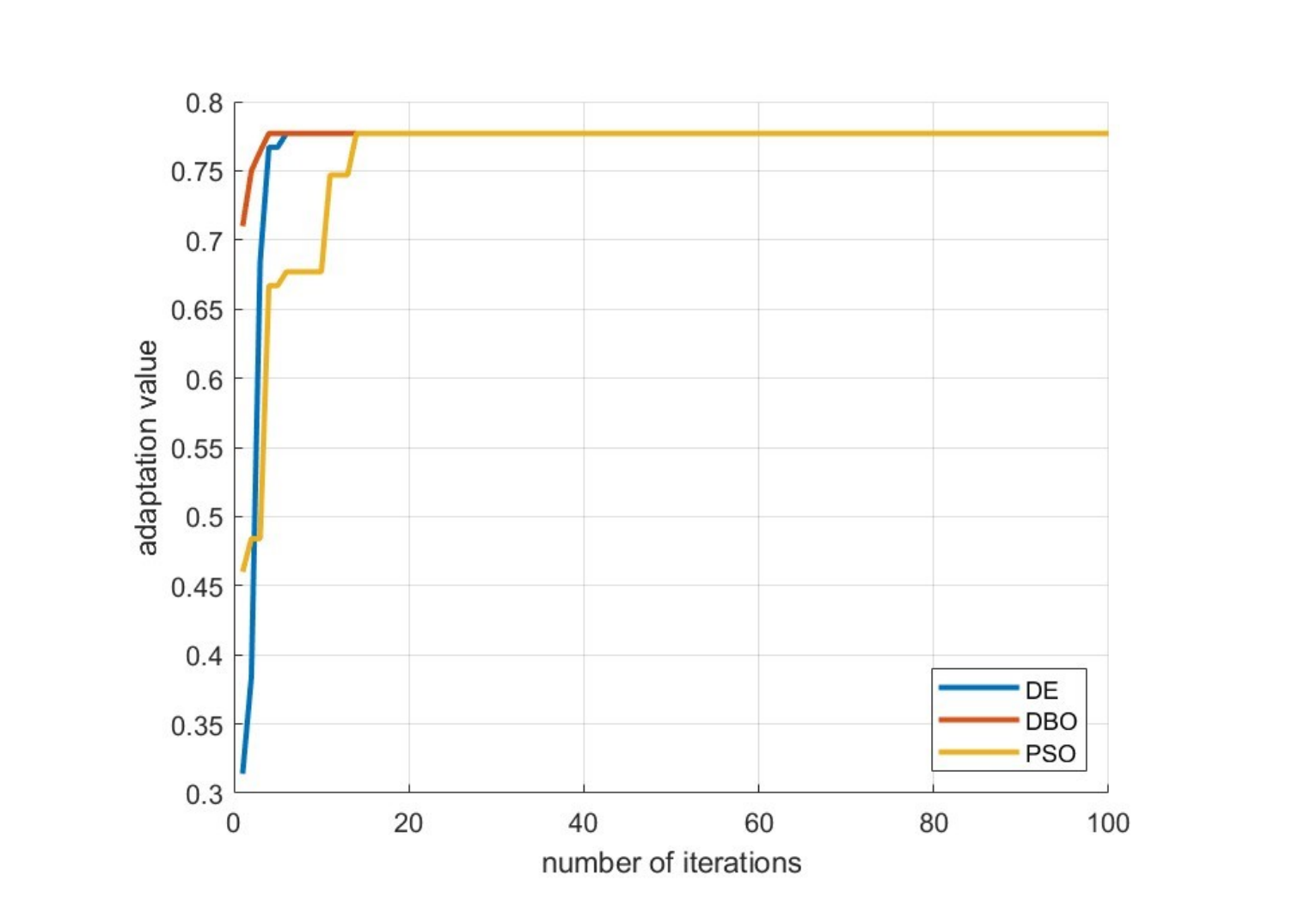} }
  \caption{Shortest robustness radius $r$ for situation B.}
   \label{Fig:sf2}
\end{figure}

In order to go verify the superiority of DBO algorithm, DE algorithm and PSO algorithm are adopted to obtain robustness radius as well. As can be seen from Figure.\ref{Fig:sf2}, the shortest distance between $working \ point$ and boundary has been chosen as the fitness function of three algorithms. By simulating such algorithms, all of these algorithms can attain the same robustness radius. However, the shortest distance between $working \ point$ and boundary converges to the optimal value by DBO algorithm with the least number of iterations.

\begin{figure}
  \centering
  \centerline{\includegraphics[width=0.53\textwidth]{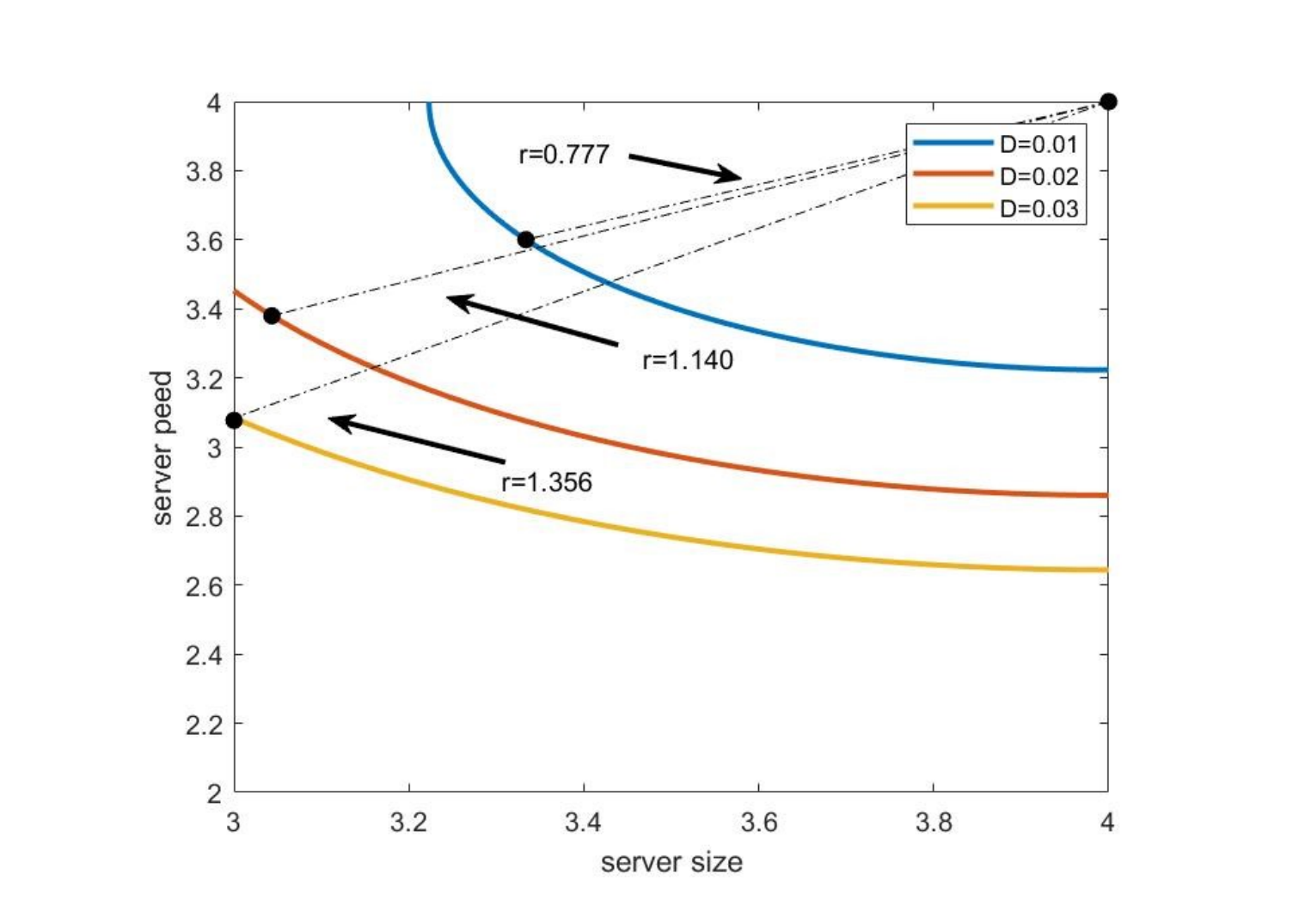} }
  \caption{Shortest robustness radius $r$ versus $D$.}
  \label{Fig:dt44}
\end{figure}

Furthermore, we perform the robustness analysis for the case presented in Figure.\ref{Fig:bj} (b) with deadline being setting to $D = 0.01$, $D = 0.02$ and $D = 0.03$, respectively. By simulating the proposed algorithm, the optimal $working \ point$ can be obtained at the same position for three cases, where the size and speed of servers are $m_z = 4$ and $s_z = 4$. As can be seen from Figure.\ref{Fig:dt44}, the robustness radius for $D = 0.01$, $D = 0.02$ and $D = 0.03$ cases are $0.777$, $1.140$ and $1.356$ respectively, for the reason that the larger the deadline is, the smaller possibility the waiting time of customers will exceed deadline caused by the perturbations in the size and speed of servers in cloud environment.

\subsection{Situation C}

By summarizing the previous discussions about robustness analysis, the maximization problems of the shortest distance from $working \ point$ to the boundaries for the profit of cloud service providers and the waiting time of customers are discussed separately. However, it is naturally to notice that the longer distance for the $working \ point$ away from the boundary in $feasible \ region$ in Figure.\ref{Fig:qk1}, the waiting time spent by customers will be longer as well, which will increase the possibility that the waiting time may exceed the deadline influenced by perturbations. In other case, if the $working \ point$ away from the boundary in $feasible \ region$ in Figure.\ref{Fig:qk3} is longer, the less profit will be earned by cloud service providers, which will increase the possibility that the profit may decline to the value lower than the acceptable lowest profit influenced by perturbations. Therefore, the requirements for cloud service providers and customers cannot be satisfied simultaneously in these cases.

\begin{figure}
  \centering
   \centerline{\includegraphics[width=0.53\textwidth]{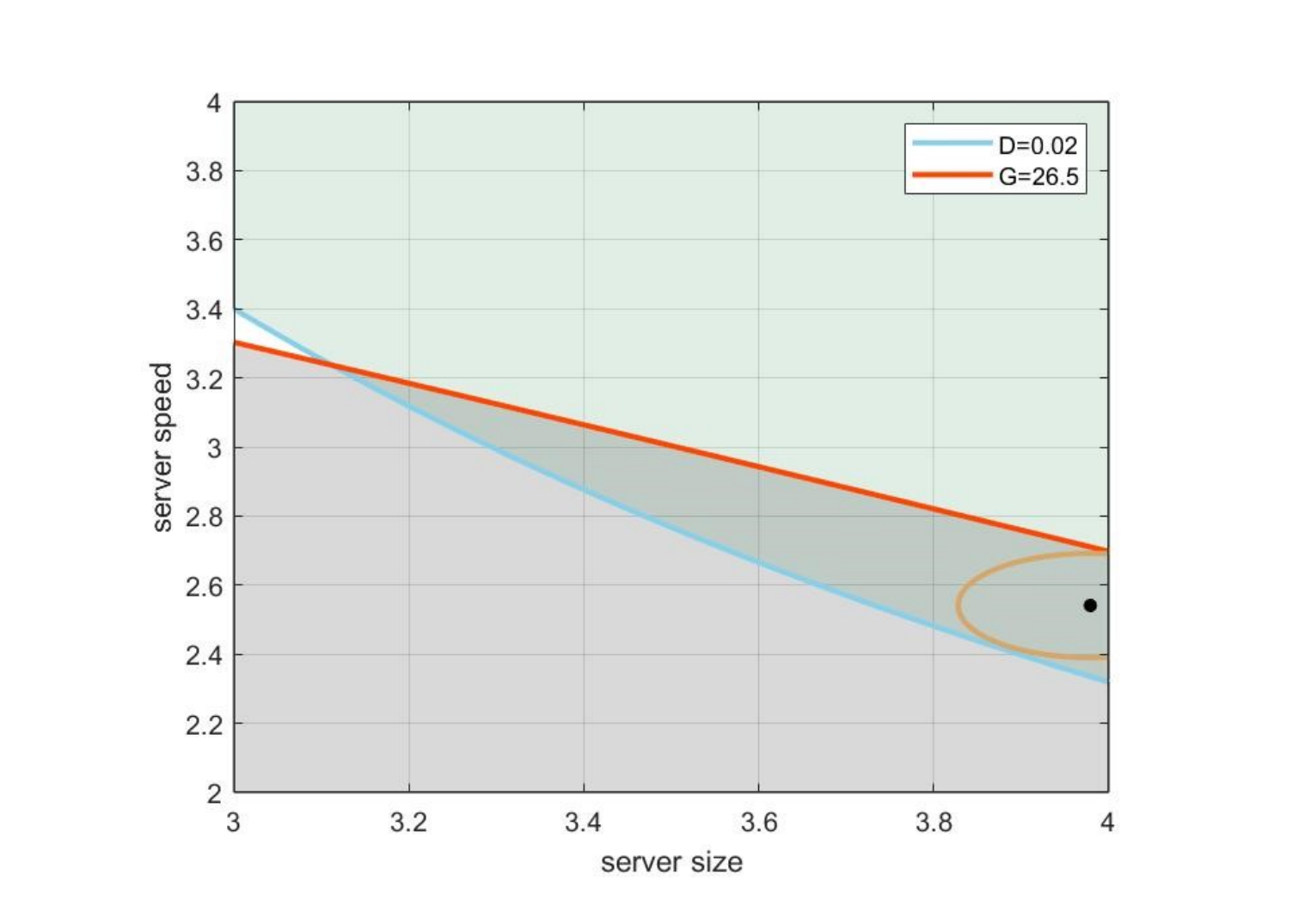} }
  \caption{The $feasible \ region$ formed by $G$ and $D$.}
  \label{Fig:qk2}
\end{figure}

On the basis of previous discussions, we try to perform robustness analysis both for the profit of cloud service providers and the waiting time of customers simultaneously, which corresponds to the cases shown in Figure.\ref{Fig:bj} (c) and (d). As can be seen from Figure.\ref{Fig:qk2}, the overlap between the region above the blue curve and the region below the red curve is considered as $feasible \ region$.

\subsubsection{Optimal size and speed}

Without loss of generality, the acceptable lowest profit of cloud service providers and deadline of customers are setting to $G = 26.5$ and $D = 0.02$ respectively. Notice that, if we blindly keep the optimal $working \ point$ away from the boundary for the profit of cloud service providers as far as possible, the waiting time of customers will easily exceed deadline influenced by perturbations. Similarly, if we blindly keep the optimal $working \ point$ away from the boundary for the waiting time of customers as far as possible, the profit of cloud service providers will easily decline to the value lower than the acceptable lowest profit influenced by perturbations. Therefore, it is reasonable to ensure the equality of the robustness radius for the profit of cloud service providers and the one for the waiting time of customers, such that the robustness requirements against the perturbations can be satisfied both for cloud service providers and customers simultaneously.

When refer to the problem for the equality of robustness radiuses, a constrained multi-objective programming model is built, which can be shown in \eqref{eqs:eqs111}. Notice that, $r_1$ and $r_2$ denote the shortest distance from a specific $working \ point$ to the boundaries with respect to profit and waiting time respectively, $G$ denotes the acceptable lowest profit earned by cloud service provider, and $D$ denotes the tolerable longest waiting time for customers. In this case, we design two objective functions, such that $r_1$ and $r_2$ can be both maximized on the basis of ensuring their equality. Then, the constraints in \eqref{eqs:eqs111} demonstrate that the $working \ point$ can only be selected in $feasible \ region$ surrounded by the boundaries.

\begin{equation}\label{eqs:eqs111}
\begin{array}{l}
    \mathop {\min }\limits_{{m_z},{s_z}} {h_1} = \left| {{r_1} - {r_2}} \right|\\
    \mathop {\max }\limits_{{m_z},{s_z}} {h_2} = {r_1} + {r_2}\\
    s.t.\left\{ \begin{array}{l}
    g\left( {{m_z},{s_z}} \right) > G\\
    f\left( {{m_z},{s_z}} \right) < D\\
    {m_z} \in \left[ {3,4} \right],{s_z} \in \left[ {2,4} \right]
    \end{array} \right.
\end{array}
\end{equation}

To solve the multi-objective programming model presented in \eqref{eqs:eqs111}, Non-dominated Sorting Genetic Algorithms II (NSGA-II) is introduced in this paper. The pseudo code for such algorithm is shown in Algorithm.\ref{alg3}. 
\begin{algorithm}[t]
    \caption{Robust scheduling based on NSGA-II algorithm}
    \label{alg3}
    \LinesNumbered
    \DontPrintSemicolon

    \KwIn{ interval of server size$ [m_{min} , m_{max}]$, server speed [$s_{min} , s_{max} ]$}
         
\KwOut{optimal $h_1,h_2$ }
\Begin{
               $t \leftarrow 1, i \leftarrow 1$;\;
               Initialize the population in the given interval;\;
               The fitness value $Y$ is calculated through Algorithm 1;\;
               $T_1 \leftarrow$ perform non-dominated sorting strategy;\;
               Sort $f$ by crowding distance for each rank ;\;
               
         \While{$t <$ Max number of iterations}{
                $P_t \leftarrow $ create parent by $T_1$ using tournament selected;\;
                $Q_t \leftarrow$ create offspring by $P_t$ using selection, crossover and mutation;\;
               $R_t \leftarrow P_t \cup Q_t$ ;\;
               The fitness value $Y_t$ is calculated through Algorithm 1;\;
               $F_t \leftarrow$ calculate all non-dominated fronts of $R_t$;\;
               $P_{t+1} \leftarrow \emptyset $ ;\;
               \While{$\left| {{P_{{\rm{t + 1}}}}} \right| \cup \left| {F_{\rm{t}}^i} \right| \le N$}{
                ${F_{\rm{t}}^i} \leftarrow $ select $i$th non-dominated front in $F_t$ using crowding distance sorting strategy;\;
                $P_t \leftarrow P_{t+1} \cup {F_{\rm{t}}^i}$;\;
                $i \leftarrow i+1$;\;

               }
              $T_{t+1} \leftarrow {P_{{\rm{t + 1}}}} \cup {F_t}\left[ {1:\left( {N - \left| {{P_{t + 1}}} \right|} \right)} \right]$;\;
              $t \leftarrow t+1$;\;

           }
       
       }
\end{algorithm}

\begin{figure}
  \centering
  \centerline{\includegraphics[width=0.53\textwidth]{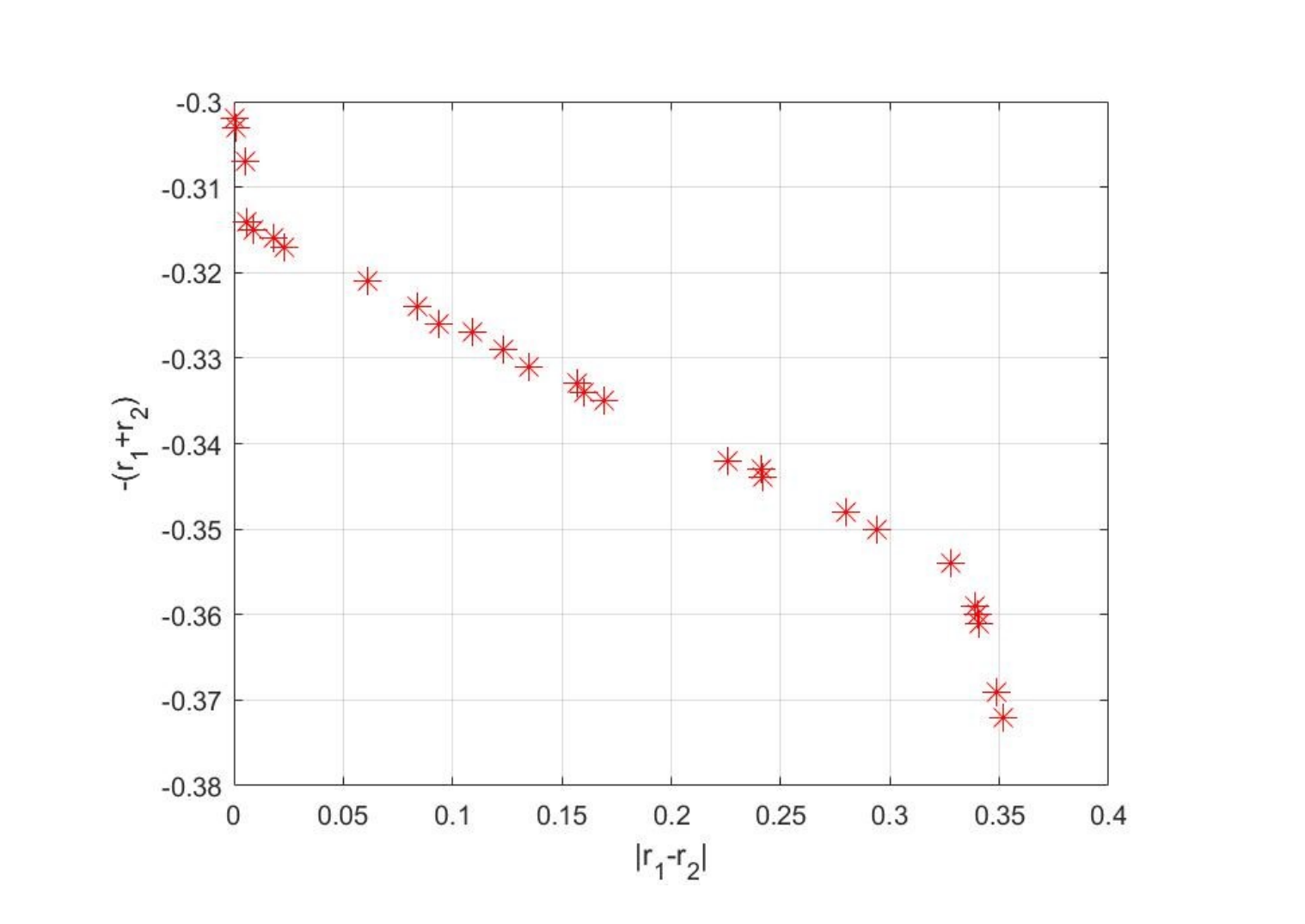} }
  \caption{Pareto frontier of $h_1$ and $h_2$ .}
  \label{Fig:pareto}
\end{figure}

By applying Algorithm.\ref{alg3} in \eqref{eqs:eqs111}, the result can be shown in Figure.\ref{Fig:pareto}. As can be seen from Figure.\ref{Fig:pareto}, each point on the pareto frontier represents an optimal solution for the multi-objective programming model. Without loss of generality, we choose the point on the left side as the optimal $working \ point$. On this basis, when we select server size and speed as $m_z = 3.979$ and $s_z = 2.541$, the shortest distances from such point to the boundaries are $r_1 = 0.151$ and $r_2 = 0.151$ respectively.

\section{Conclusion}\label{sec: conclusion}

In this paper, we investigate the robust scheduling problem in the presence of some unpredictable perturbations in cloud computing system. First, a cloud computing system is built based on queuing theory, in which profit and waiting time are declared as the targets of cloud service providers and customers respectively. Then, to keep the profit earned by cloud service providers away from their lowest acceptable profit, and keep the waiting time spent by customers away from their longest acceptable waiting time as much as possible, we define a robustness metric method, and then propose heuristic optimization algorithm to obtain the optimal configuration of the number and speed of servers. The superiority of the proposed algorithm is validated by comparing with DE and PSO algorithms.


%

\section*{Acknowledgment}
The authors would like to thank the anonymous reviewers for their valuable comments and suggestions.
This work has been supported by Excellent Young Scientists Fund of Hunan Provincial  (Grant Nos. 22B0156) Guangdong Basic and Applied Basic Research Foundation (Grant No. 2021A1515110383) and Foshan Shunde Key and Technology Projects (Grant No.213021800254) 

\bibliographystyle{IEEEtran}
\bibliography{IEEEabrv,reference}




\end{document}